\documentclass[usenatbib]{mn2e}
\usepackage{times}
\usepackage{graphicx} 
\usepackage{amsmath}
\usepackage{mathabx}
\usepackage{natbib}

\def \fc {$f_{\rm convert}$}
\begin{document}

\title[Satellite Quenching Evolution]{Evidence for a change in the dominant satellite galaxy quenching mechanism at $z=1$}
\author[Balogh et al.]{Michael L. Balogh$^{1}$, Sean L. McGee$^{2}$,  Angus Mok$^{3}$, Adam Muzzin$^{4}$, 
\newauthor Remco F.J. van der Burg$^{5}$,  Richard G. Bower$^{6}$, Alexis Finoguenov$^{7}$,
\newauthor Henk Hoekstra$^{8}$, Chris Lidman$^{9}$, John~S. Mulchaey$^{10}$, Allison Noble$^{11}$, Laura C. Parker$^{3}$
\newauthor Masayuki Tanaka$^{12}$, David J. Wilman$^{13,14}$, Tracy Webb$^{15}$, Gillian Wilson$^{16}$, Howard K.C. Yee$^{11}$
\\
$^{1}$Department of Physics and Astronomy, University of Waterloo, Waterloo, Ontario, N2L 3G1, Canada\\
$^{2}$School of Physics and Astronomy, The University of Birmingham, Birmingham B15 2TT, UK\\
$^{3}$Department of Physics and Astronomy, McMaster University, Hamilton, Ontario, L8S 4M1 Canada\\
$^{4}$Institute of Astronomy, University of Cambridge, Madingley Rd, Cambridge CB3 0HA, UK\\
$^{5}$Laboratoire AIM, IRFU/Service d'Astrophysique - CEA/DSM - CNRS - Université Paris Diderot, Bat. 709, CEA-Saclay, 91191 Gif-sur-Yvette Cedex, France\\
$^{6}$Department of Physics, University of Durham, Durham, UK, DH1 3LE\\
7$^{7}$Department of Physics, University of Helsinki, Gustaf H{\" a}llstr{\" o}min katu 2a, FI-00014 Helsinki, Finland\\
$^{8}$Leiden Observatory, Leiden University, PO Box 9513, 2300 RA Leiden, The Netherlands\\
$^{9}$Australian Astronomical Observatory, 105 Delhi Rd, North Ryde NSW Australia 2113\\
$^{10}$Observatories of the Carnegie Institution, 813 Santa Barbara Street, Pasadena, California, USA\\
$^{11}$Department of Astronomy and Astrophysics, University of Toronto, 50 St. George Street, Toronto, Ontario, Canada M5S 3H4\\
$^{12}$National Astronomical Observatory of Japan, 2-21-1 Osawa, Mitaka, Tokyo 181-8588, Japan\\
$^{13}$Universit{\" a}ts-Sternwarte M{\" u}nchen, Scheinerstrasse 1, D-81679 M{\" u}nchen, Germany\\
$^{14}$Max--Planck--Institut f{\" u}r extraterrestrische Physik, Giessenbachstrasse 85748 Garching Germany\\
$^{15}$Department of Physics, 3600 rue University, Montreal, QC, Canada H3A 2T8\\
$^{16}$Department of Physics \& Astronomy, University of California, Riverside, 900 University Avenue, Riverside, CA 92521\\
}
\date{\today}
\maketitle
\begin{abstract}
We present an analysis of galaxies in groups and clusters at $0.8<z<1.2$, from the GCLASS and GEEC2 spectroscopic surveys.   We compute a ``conversion fraction'' \fc\ that represents the fraction of galaxies that were prematurely quenched by their environment.  For massive galaxies, $M_{\rm star}>10^{10.3}M_\odot$, we find \fc$\sim 0.4$ in the groups and $\sim 0.6$ in the clusters, similar to comparable measurements at $z=0$.  This means the time between first accretion into a more massive halo and final star formation quenching is $t_p\sim 2$ Gyr. This is substantially longer than the estimated time required for a galaxy's star formation rate to become zero once it starts to decline, suggesting there is a long delay time during which little differential evolution occurs.   In contrast with local observations we find evidence that this delay timescale may depend on stellar mass, with $t_p$ approaching $t_{\rm Hubble}$ for $M_{\rm star}\sim 10^{9.5}M_\odot$.  The result suggests that the delay time must not only be much shorter than it is today, but may also depend on stellar mass in a way that is not consistent with a simple evolution in proportion to the dynamical time.  Instead, we find the data are well-matched by a model in which the decline in star formation is due to ``overconsumption'', the exhaustion of a gas reservoir through star formation and expulsion via modest outflows in the absence of cosmological accretion.  Dynamical gas removal processes, which are likely dominant in quenching newly accreted satellites today, may play only a secondary role at $z=1$. 
\end{abstract} 

\section{Introduction}
Observations of galaxies with sufficiently precise redshifts and stellar mass estimates, coupled with cosmological dark matter simulations, have led to the development of an increasingly clear empirical description of massive galaxy evolution.  In particular, it has been established that galaxy formation is most efficient within haloes of $\sim 10^{12}M_\odot$ \citep[e.g.][]{Leauthaud12}.  The baryon accretion rate, generally assumed to trace the dark matter accretion rate, is largely decoupled from the stellar mass growth, especially at late times \citep{Behroozi13}.  Hydrodynamic and ab-initio semi-analytic models, which attempt to incorporate as many physical processes as possible, are also becoming impressively accurate \citep[e.g.][]{Guo,Henriques,Ill_Genel,EAGLE}, but several stubborn problems persist.  One is the rapid decline in global star formation rate (SFR) since $z\sim 2$, and its dependence on stellar mass \citep{Bower12,WWN,Weinmann11,DL12,Ill_Genel,Hen15,Furlong,Barre15}.  This problem reflects the fact that this evolution is very different from the total mass accretion rate over this time.  Decoupling the baryonic processes from the dark matter growth requires substantial feedback and mass ejection \citep[e.g.][]{McCarthy11}; this is poorly constrained observationally and is generally modeled with simple, parametric descriptions \citep[e.g.][]{Hen15}.   

Another persistent problem is in the prediction of satellite galaxy properties; models consistently predict satellite populations that ceased star formation prematurely, compared with observations \citep[e.g.][]{Weinmann11,V+14}.  Attempts to solve this problem with a more physical treatment of gas stripping \citep[e.g.][]{McCarthy_rps} generally fail to reproduce the observed star-formation rate (SFR) distribution of these galaxies \citep[e.g.][]{Font,Weinmann10}.  

Recently, \citet{overconsumption} suggested that the two problems might be related, and that the overquenching of satellite galaxies indicates that the feedback and outflows thought to be required to drive the global evolution in SFR are too strong.  In the presence of strong outflows, and in the absence of cosmological accretion, the timescale for satellite galaxies to consume their gas can be much shorter than the time for the gas to be stripped through dynamical processes. This will lead to a very different form of satellite galaxy evolution than predicted by gas stripping models, for example. An alternative, proposed explanation for the discrepancy with simple models is the assembly bias of dark matter haloes \citep[e.g.][]{AB,Zentner}.  Haloes of a given mass do not all have the same assembly history, and it has been shown that associating the oldest haloes with the oldest galaxies leads to an improved description of satellite galaxy properties at $z=0$ \citep{Watson14}.  

Most of the best constraints on these and other models come from large spectroscopic surveys at low redshift.  Measurements of the redshift evolution of galaxies in different environments have potential to break the degeneracy between various models.  This is because the rate of evolution in galaxy properties (star formation rate, gas fractions, etc.) is different from that of halo masses, and of the relative importance of assembly bias.  However, despite many years of studying galaxy evolution in groups and clusters \citep[e.g.][]{BO78a}, this remains a formidable challenge, because of the difficulty of the measurements and the dominant role played by systematic uncertainties.

The evolution of massive, central galaxies is now fairly well constrained from large spectroscopic and photometric surveys \citep[e.g.][]{CFRS,Noeske,COSMOS-Karim,VIPERS-D,Ultravista,CANDELS-Lee}.  From these same surveys it is possible to learn something about the evolving role of environment, from measurements of local density or group catalogues \citep[e.g.][]{Giodini,George2, Knobel_qf, Panstarrs-lin}.  In the context of the halo model \citep[e.g.][]{CW09}, though, the most important characteristic of the environment is whether or not a galaxy is a satellite within a more massive halo \citep[e.g.][]{Woo}, and a clear picture of this can only be obtained from spatially complete, deep spectroscopic surveys.  

The most massive clusters have long been studied in this context \citep{BO,CNOC97,Ellingson,Andreon10}, and observations of these systems extend out well beyond $z=1$  \citep[e.g.][]{Demarco07,Fassbender11,Andreon14,GCLASS,madcows14}.  Comparable work on more common, lower-mass haloes at $z>0.3$ --- the progenitors of today's massive clusters --- is more difficult and hence more limited, but several studies have shown that group galaxies indeed evolve differently from the field, at least for $z<0.8$ \citep[e.g.][]{GEEC1,GEEC1_MstarMhalo,Pogg08,R09,McGee11,Knobel_qf}.  
At $z>0.8$ however, little is known about galaxy evolution in group and low--mass cluster haloes.  The best spectroscopic data available are from the GCLASS survey of ten galaxy clusters at $0.8<z<1.2$ \citep{GCLASS}, and the GEEC2 survey of lower mass groups at a similar redshift \citep{GEEC2}.  In a series of papers these surveys have been used to explore the correlations between stellar- and halo mass \citep{GEEC2,Muzzin-ps,RvdB,RvdB1}, and the environmentally--driven quenching of satellite galaxies \citep{GCLASS,Mok1,Mok2}.  In this paper we combine both samples for a homogeneous analysis of central and satellite galaxies, spanning almost two orders of magnitude in halo mass at $0.8<z<1.2$.  

Unless otherwise stated, throughout this paper we assume a WMAP9 \citep{WMAP9} cosmology ($H_\circ$=69.3km/s, $\Omega_m=0.286$, $\Lambda=0.713$).  All magnitudes are on the AB system.  Halo masses and sizes are generally characterised by the radius within which the average mass density is 200 times the critical density of the Universe at the redshift of the cluster, $R_{200}$.

\section{Data}
\subsection{GEEC2} 
GEEC2 is a spectroscopic survey of galaxies in $11$ groups within the COSMOS field. It consists of 603 galaxies with secure redshifts, 162 of which are group members.  Details of the target selection and spectroscopic observations have been  thoroughly described in \citet{geec2_data}.  Candidate group targets were selected from an X-ray selected catalogue that was later published in \citet{George2,George1}.  Spectroscopy was obtained using the GMOS spectrograph on Gemini-South, over two semesters. We used the nod-and-shuffle feature, with the R600 grating and 1x3\arcsec\ slits.  The analysis in this paper also incorporates the DR2 release of the 10K zCOSMOS spectroscopic survey \citep{zCOSMOS}.

GMOS spectroscopic targets were selected based on their $r-$band magnitude, from the deep photometric catalogues of \citet{Capak}, and photometric redshifts from \citet{Ilbert}.  The photometric redshift catalogue we use has no explicit magnitude cut.  The relevant magnitude limits for our purposes are the 80 per cent completeness limit in the detection catalogue, $i=26.5$ \citep{Capak}, and the 5$\sigma$ limiting magnitude of the IRAC [3.6]$\mu$m catalogue, AB$<24.0$ \citep{Sanders}.

\begin{figure}
{\includegraphics[clip=true,trim=0mm 0mm 0mm 0mm,width=3.5in,angle=0]{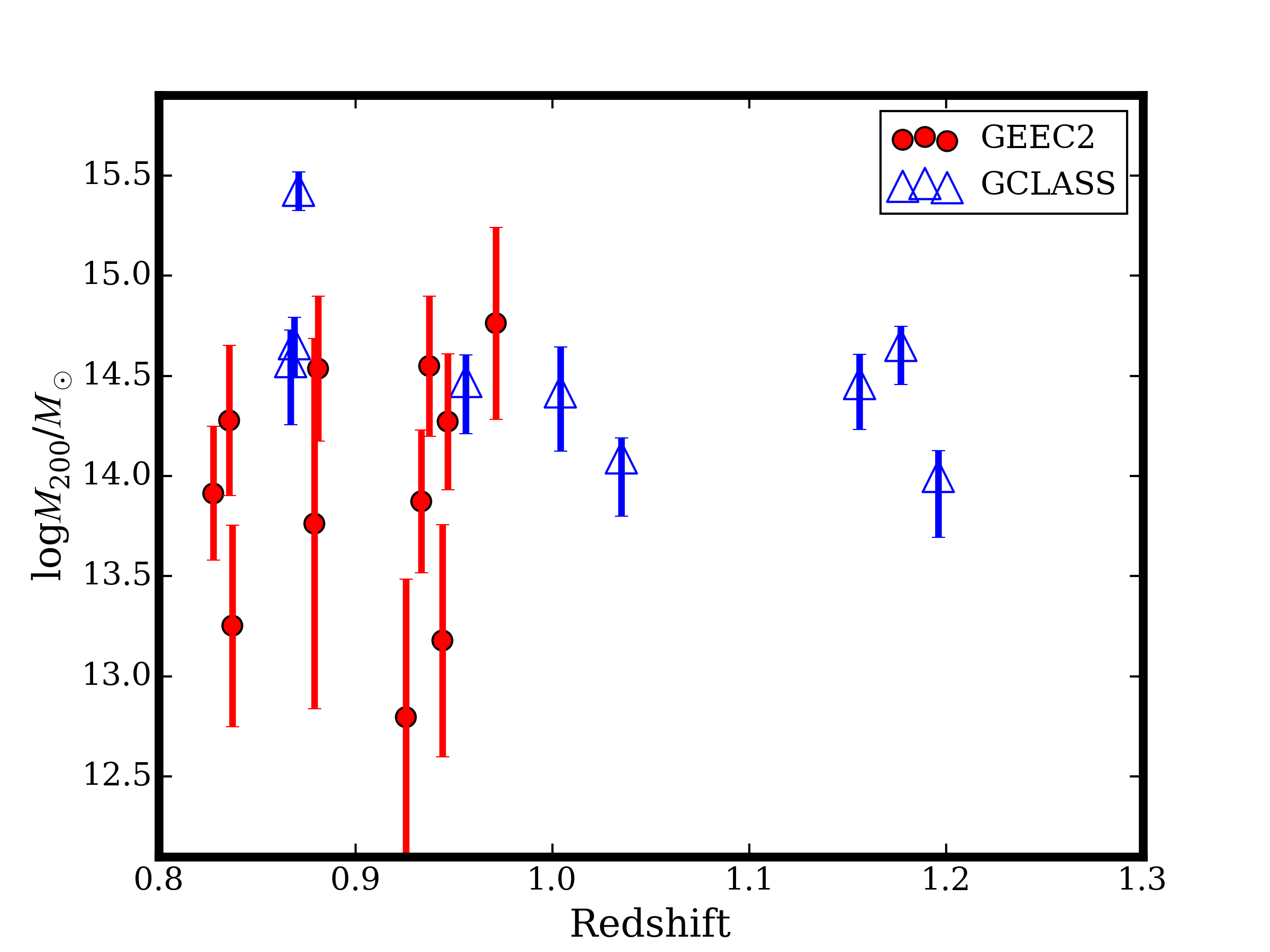}}
        \caption{The dynamical mass is shown as a function of system redshift for all groups and clusters in the GEEC2 (red circles) and GCLASS (blue triangles) samples.}
\label{fig:mz}
\end{figure}

Halo masses are computed from the velocity dispersions, within $R_{200}$ and within $R_{rms}$, the {\it rms} position of all spectroscopic members.  As argued in \citet{geec2_data} we choose the larger of these two radii, $R_{max}$, as some poorly populated systems have few members within $R_{200}$.  The corresponding dynamical mass within this radius shown, as a function of group redshift, with red circles in Figure~\ref{fig:mz}.

For galaxies without spectroscopic redshifts, we treat their membership statistically, using a probability $p_g$ determined from the photometric redshift probability distribution function.  This procedure is described in detail in \citet{geec2_data}, and includes a correction for the bias that results from targeting known overdensities. To summarize, we first compute, from the distribution function, the probability $p(z_{\rm group},3\sigma)$ that a galaxy lies within 3$\sigma$ of a group redshift, $z_{\rm group}$, where $\sigma$ is the group velocity dispersion.  By comparing this quantity with the fraction of spectroscopic members within $3\sigma$, we determine that the actual probability of a galaxy being in the group is larger, and given by $p_g=p(z_{\rm group},3\sigma)^{0.5}$. This neglects any dependence of the correction on galaxy colour, stellar mass or apparent magnitude. Galaxies without spectroscopy that are assigned to a given group in this way are assumed to be at exactly the group redshift, for the purposes of computing stellar masses and rest-frame colours.  In Appendix~\ref{sec-app1} we examine the sensitivity of our conclusions to the systematic uncertainties associated with photometric redshifts.  

Stellar masses for all galaxies with spectra were computed as described in \citet{Mok1}, based upon \citet{BC03} templates \citep[using the more recent models of ][]{CB07} with a \citet{Chabrier} initial mass function.  For the photometric sample, stellar masses are computed from a fit to the [3.6]$\mu$m fluxes and $(r-i)$ colours of the spectroscopic sample, as described in \citet{geec2_data}.  The stellar mass limit for passive galaxies at the IRAC 5$\sigma$ limit AB=24.0 and our maximum redshift of $z=1$ corresponds to $10^{9.5}M_\odot$, though we caution that some incompleteness may set in at $M_{\rm star}<10^{10}M_\odot$ as a small fraction of these galaxies will be undetected in $i$.

Galaxies are classified as star-forming, passive or intermediate, based on their location in an optical--IR colour plane, as described in \citet{Mok1} and \citet{Mok2}.   Throughout this paper we consider the intermediate--type galaxy as part of the passive population.  This classification relies on measurements in $V$, $z$, $J$, and IRAC [3.6]$\mu$m, and allows the distinction between dusty, star-forming galaxies and truly passive galaxies.  For this paper, we require detections in the three reddest bands.  Galaxies that are undetected in $V$ are classified as passive. This may lead to an overestimate of the passive galaxy fraction at the lowest stellar masses considered here. The main conclusions of this paper derive from an observation that the fraction of passive, low-mass galaxies is surprisingly {\it low} compared with naive expectations; thus our treatment of these non-detections in $V$ is appropriately conservative, as any other assumption would make our conclusions even stronger.

Although the groups were selected in part due to the presence of X--ray emission, three of them are unlikely to be associated with that emission \citep{geec2_data}.  Group 213a was a serendipitous discovery behind group 213, and the distribution of spectroscopic members in groups 121 and 161 is significantly offset from the centroid of the nearby X--ray emission.  It is likely that the initial estimated redshift of the groups corresponding to that emission, based on spectroscopy of only a few galaxies, was incorrect.  Since the spectroscopic follow-up preselected targets to lie at that predicted redshift, the GEEC2 spectroscopy is not sufficient to confirm the presence of a foreground or background group as the more likely source.  We include these groups in our analysis but have verified that none of our conclusions are significantly impacted if we exclude them.  This is unsurprising, as the groups have few members and we do not use the X--ray emission in our analysis.

\subsection{GCLASS}\label{sec-gclass}
GCLASS \citep{GCLASS,RvdB} is a sample of ten galaxy clusters at $0.85<z<1.35$, selected from the 42 deg$^2$ SpARCS survey \citep{Sparcs-muzzin,Sparcs-wilson,Demarco10}.  SpARCS makes use of deep $z$-band imaging in the SWIRE \citep{SWIRE} fields, to identify overdensities in $z-[3.6]$ colour.  Ten of these systems were chosen for follow-up with GMOS in nod-and-shuffle mode, using the R150 grating and $1\times3$ arcsecond slits.  Masks were designed to maximize efficiency in the rich cluster cores, and clusters were generally observed with between 4 and 6 MOS masks, with 3 hour exposure times each.  Most of the GCLASS data and analysis comes from the 9 clusters at $0.85<z<1.25$; the highest redshift system only contributes 20 members, with limited wavelength coverage.   The full sample comprises 457 cluster members, obtained in 222 hours of Gemini time over three years.  Analysis of the brightest cluster galaxies in GCLASS is presented in \citet{Lidman12,Lidman13}, while the dynamics of the cluster population are considered in papers by \citet{GCLASS-Noble} and \citet{Muzzin-ps}.  The dynamical masses and redshifts of the clusters are shown as blue triangles in Figure~\ref{fig:mz}.

Stellar masses are computed as described in \citet{RvdB1}, and are derived from the {\sc FAST} SED-fitting code \citep{FAST}, using \citet{BC03} models and assuming a \citet{Chabrier} initial mass function and smooth, exponentially--declining star formation histories.  \citet{RvdB1} showed that the stellar masses derived in this way were in good agreement with those from COSMOS.  The difference between the \citet{BC03} models and the \citet{CB07} models used for GEEC2 is generally small at these redshifts \citep[e.g.][]{Ilbert10}.  The use of $\tau-$models, and the limited photometry compared with that available for GEEC2, may also lead to small systematic differences in mass estimates.  On average, GEEC2 masses are large than those in the Ultravista catalogue by 0.07 dex.  This difference is small relative to our statistical uncertainties, and we neglect this difference, making no correction.

The stellar mass functions, velocity dispersions, and stellar mass profiles can be found in \citet{RvdB1} and \citet{RvdB}.  These make use of background-subtracted photometric redshifts to supplement and somewhat extend the depth of the available spectroscopy.  Galaxy classification is based on location in the UVJ colour--colour diagram, in a similar way to GEEC2, but without the intermediate classification or rest-NUV information.  The stellar mass completeness limit is calculated separately for each cluster, and is derived from the 80 percent completeness depth of the $K$-band imaging, and the highest M/L ratio measured for galaxies near that limiting magnitude.  The exposure times were adjusted to achieve an approximately uniform mass completeness across the redshift range of the sample.  The stellar mass limits range from $9.92<\log{M/M_\odot}<10.53$.  
We use their mass functions as published\footnote{Note that these mass functions are based on a slightly different cosmology from that used in the present paper, $\Omega_m=0.3$ and $H_\circ=70$km/s/Mpc.  This makes a $<3$ per cent difference to luminosity-related quantities like stellar mass.} for the analysis in the present paper.  When computing stellar mass functions, only the clusters with sufficiently deep data are included in a given mass bin.  There are six clusters with stellar mass limits $\log{M/M_\odot}\leq 10.15$, and these are the only ones that contribute to every bin in the mass function. In addition, a correction of up to 37 per cent is applied to the lowest three stellar mass bins, separately for star--forming and passive galaxies, to account for residual incompleteness inferred from a comparison with UltraVISTA photometry.

\subsection{Sloan Digital Sky Survey Reference Sample at $z=0$}
For comparison we use the seventh data release of the Sloan Digital Sky Survey \citep[SDSS,][]{SDSS}.  In particular we 
use the compilation and definitions of \citet{Omand}.  This compilation is based on the group catalogue of \citet{YangGC}, and data drawn from catalogs made by \citet{Blanton2005}, \citet{Brinchmann}, \citet{Simard}, and \citet{YangDR7}.  We define the most luminous galaxy in each group to be the central galaxy and other members to be satellites. Isolated galaxies (those not linked to any group) are also defined as centrals.  The definition of passive galaxies is based on the bimodality of the specific SFR distribution as a function of stellar mass.  Since star-forming galaxies are quite clearly separated from passively-evolving galaxies, both in colour-colour space and in their derived SFR distribution, our results are insensitive to the details of precisely how the dividing line is drawn, or whether colours or inferred SFR are used as the distinguishing parameter.

\section{Passively evolving galaxies in groups and clusters at $z=1$}
\begin{figure}
{\includegraphics[clip=true,trim=0mm 0mm 0mm 0mm,width=3.5in,angle=0]{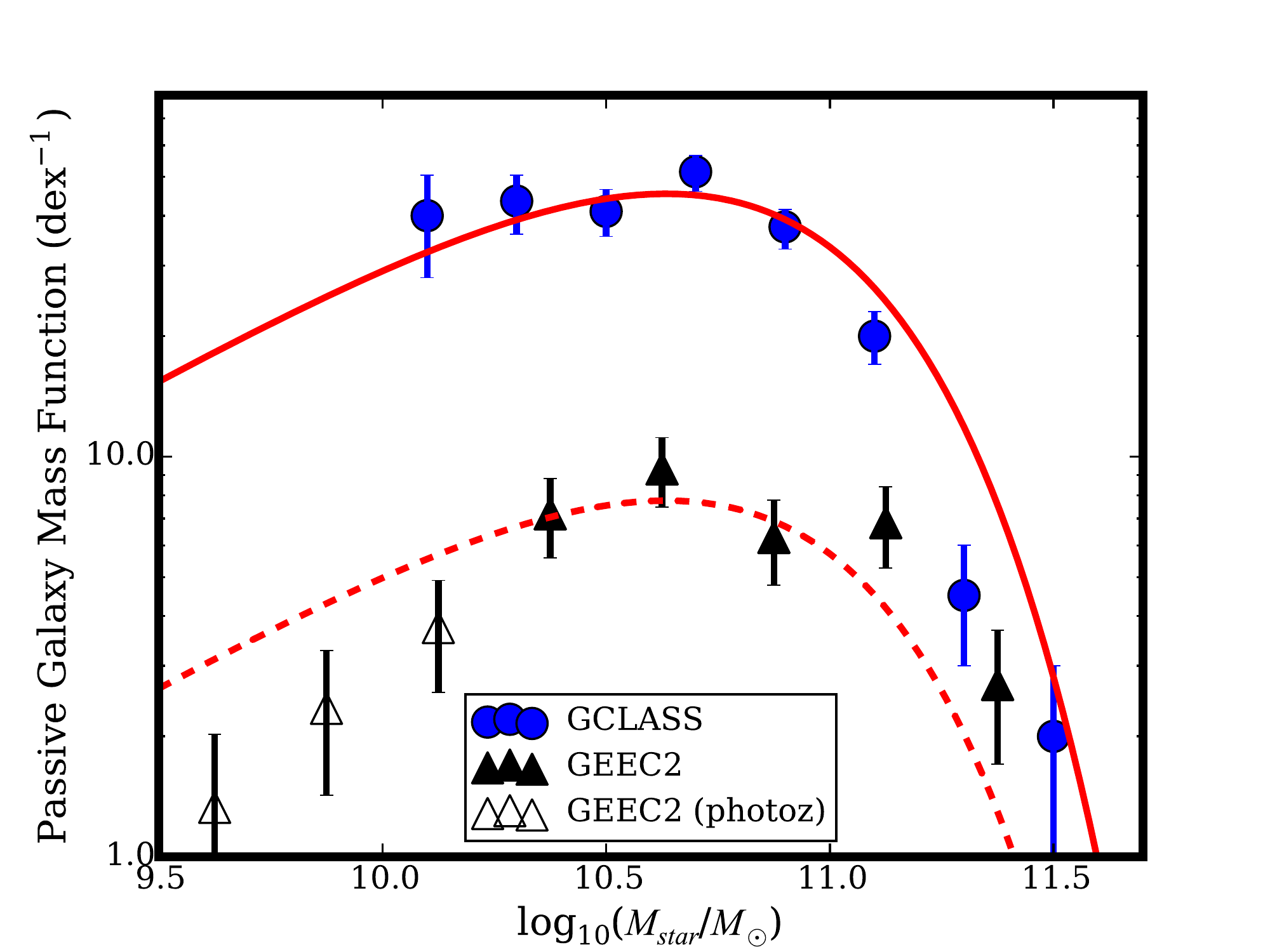}}
        \caption{The mass function of passive galaxies is shown for the GEEC2 and GCLASS samples.  The normalization reflects the average number of galaxies per system in each sample, and reflects the fact that GCLASS clusters are an order of magnitude more massive than the GEEC2 groups.  For comparison we show the mass function of passive galaxies in the field, from Ultravista \citep{Ultravista}, as the red lines, arbitrarily renormalized to  match the integrated stellar mass in galaxies within each of the GEEC2 (dotted line) and GCLASS samples (solid line).}
\label{fig:pmf}
\end{figure}
The stellar mass function of satellite galaxies in GCLASS and GEEC2 has been presented in \citet{RvdB1} and \citet{Mok1}, respectively.  We focus here on the passive galaxy mass function, and compare these two results directly in Figure~\ref{fig:pmf}.  For GEEC2 we estimate the mass function including galaxies with photometric redshifts by integrating the $p_g$ probabilities.  This approach provides results that are consistent with those of \citet{Mok1}, where a completeness correction was applied to the spectroscopic sample, but also allows us to extend that result to masses below the spectroscopic limit, shown as the open symbols. For comparison we show the mass function for passive field galaxies at $z=0.9$ from the Ultravista survey \citep[][ Schechter function parameters $\log{M^\ast/M_\odot}=10.83$, $\alpha=-0.36$]{Ultravista}, renormalized to match the integrated stellar mass in galaxies within each of the group and cluster samples. 

Both GCLASS and GEEC2 show a passive galaxy mass function shape that is in reasonable agreement with that of the field, down to the spectroscopic limit. Using the photometric redshifts in COSMOS to consider stellar masses below $10^{10.3}M_\odot$, the GEEC2 mass function declines steeply, showing even fewer passive galaxies than the field.  The drop in abundance of passive galaxies with decreasing stellar mass (or luminosity) has been seen by others in moderate redshift clusters and groups \citep[e.g.][]{DL07,Bildfell,Martinet}.  However, this remains controversial \citep[e.g.][]{Andreon08,dPPB,Andreon14}, and the shape of the group mass function presented here in particular, which implies a Schechter parameter $\alpha\sim 0$, is extreme.  We have confirmed that this difference in shape relative to the field is seen even if the whole analysis is done with photometric redshifts (using the computed values of $p_g$), ignoring the spectroscopy.  Thus it does not seem that the change in slope is due to the different redshift estimator.  Furthermore, we can consider an upper limit on the passive galaxy mass function, by assuming that all passive galaxies with $p_g>0.1$ are group members, and neglecting the bias correction in the photometric redshift probability for star-forming galaxies.  The relevant analysis in this paper is repeated with that conservative assumption, and presented in Appendix~\ref{sec-app1}.  These upper limits trace the shape of the field galaxy mass function.  Thus it appears that the lack of low-mass, passive galaxies is real, though spectroscopic confirmation is required.

\begin{figure}
{\includegraphics[clip=true,trim=0mm 0mm 0mm 0mm,width=3.5in,angle=0]{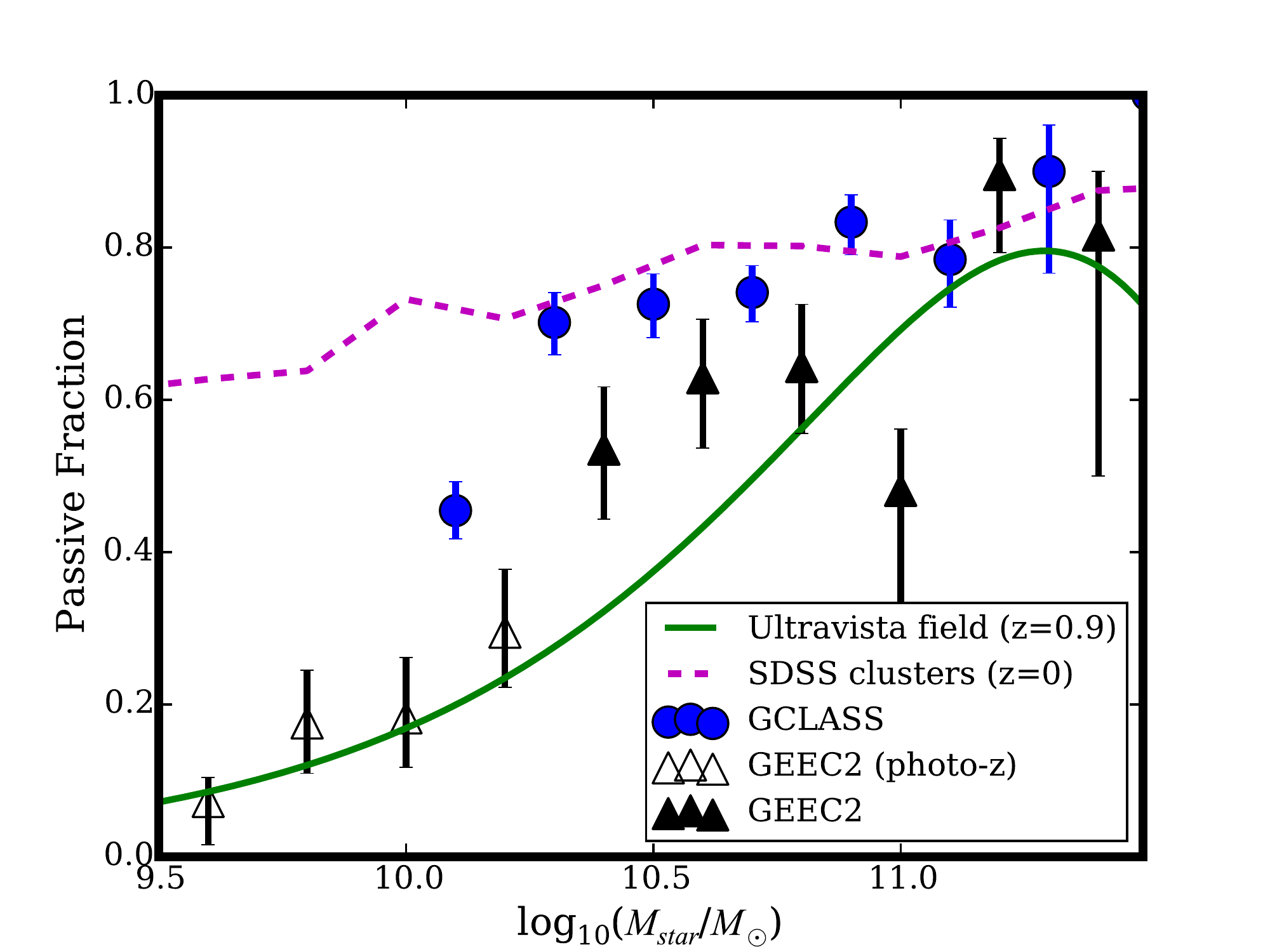}}
        \caption{The fraction of passive galaxies as a function of stellar mass is shown for the GEEC2, GCLASS samples, compared with a sample of SDSS clusters \citep{Omand} and the Ultravista \citep{Ultravista} field at $z=0.9$.  The GCLASS (cluster) sample shows high passive fractions, consistent with observations at $z=0$.  The group sample shows a more modest enhancement of passive fraction, with no significant enhancement at masses below $M_{\rm star}=10^{10.3}M_\odot$. }
\label{fig:qf}
\end{figure}

In Figure~\ref{fig:qf} we show, for the same samples, the fraction of all galaxies that are classified as passive, as a function of stellar mass.  For the clusters, in the GCLASS sample, the fraction of passive galaxies is always much higher than it is in the $z=0.9$ field from Ultravista (green, solid line), as previously shown by \citet{GCLASS} and \citet{RvdB1}.  This demonstrates that environment continues to play a significant role in determining galaxy properties at this redshift.  Interestingly, these fractions are very similar to the fractions observed in local clusters of similar halo mass ($\log{M_{\rm halo}/M_\odot}>14.2$), from \citet{Omand}, shown by the dashed magenta line.  The exception is in the lowest-mass bin, which has a significantly lower passive fraction.  As noted in \S~\ref{sec-gclass}, the lowest mass bins in this sample are computed from a subset of the full cluster sample, and are subject to a small approximate completeness correction.  While these points represent our best estimate of the measurement expected from a complete sample, it is appropriate to treat them as provisional pending confirmation from deeper data.

The group sample shows a more modest enhancement of the passive fraction relative to the surrounding field, most evident atintermediate masses $10.3<\log{M_{\rm star}/M_\odot}<11.0$.  At lower masses the passive fraction is indistinguishable from the field.   It seems a concern that this change in behaviour relative to the field occurs below the spectroscopic limit of the sample.  In Appendix~\ref{sec-app1} we explore how the result is influenced by more conservative choices in photometric redshift selection, and conclude that a passive fraction that remains equally enhanced relative to the field below $\log(M_{\rm star}/M_\odot)=10.3$ is a robust upper limit to the true fraction.  Moreover it is interesting that a drop in relative passive fraction, at the same stellar mass, is also seen in the GCLASS sample.

To explore this further we consider the ``conversion fraction'', \fc\ \citep{P14}, as first introduced by \citet{vdB}, to estimate what fraction of star--forming galaxies accreted by groups and clusters have had their star formation quenched by the environment \citep[see also][]{Wetzel13,Hirschmann}.  This is given by
\begin{equation}\label{eq-qe}
f_{\rm convert}(M_{\rm star})=\frac{f_{p,clus}(M_{\rm star})-f_{p,field}(M_{\rm star})}{1-f_{p,field}(M_{\rm star})},
\end{equation}
where $f_{p,clus}(M_{\rm star})$ and $f_{p,field}(M_{\rm star})$ are the fraction of passively evolving galaxies in the cluster (or group) and the field\footnote{For GEEC2, the field sample excludes spectroscopic members of the targeted groups.}, respectively, at stellar mass $M_{\rm star}$.  By taking the field value at the same epoch as the cluster sample, this represents the excess quenching that would occur, over and above what would happen if the galaxy remained a central. In Appendix~\ref{sec-app2} we discuss the effect of choosing $f_{p,field}(M_{\rm star})$ at the epoch of accretion, instead \citep[see also ][]{Hirschmann}.  Neither approach is demonstrably correct, but the approach taken here is relevant to discover the minimum role environment might play, by assuming that the mechanisms for quenching massive galaxies remain uninterrupted when a galaxy becomes a satellite.  Even in this case we acknowledge that this conversion fraction is a simplification; for example, our interpretation of Equation~\ref{eq-qe} neglects differential growth in stellar mass between group and field galaxies, which may be significant. 
\begin{figure}
{\includegraphics[clip=true,trim=0mm 0mm 0mm 0mm,width=3.5in,angle=0]{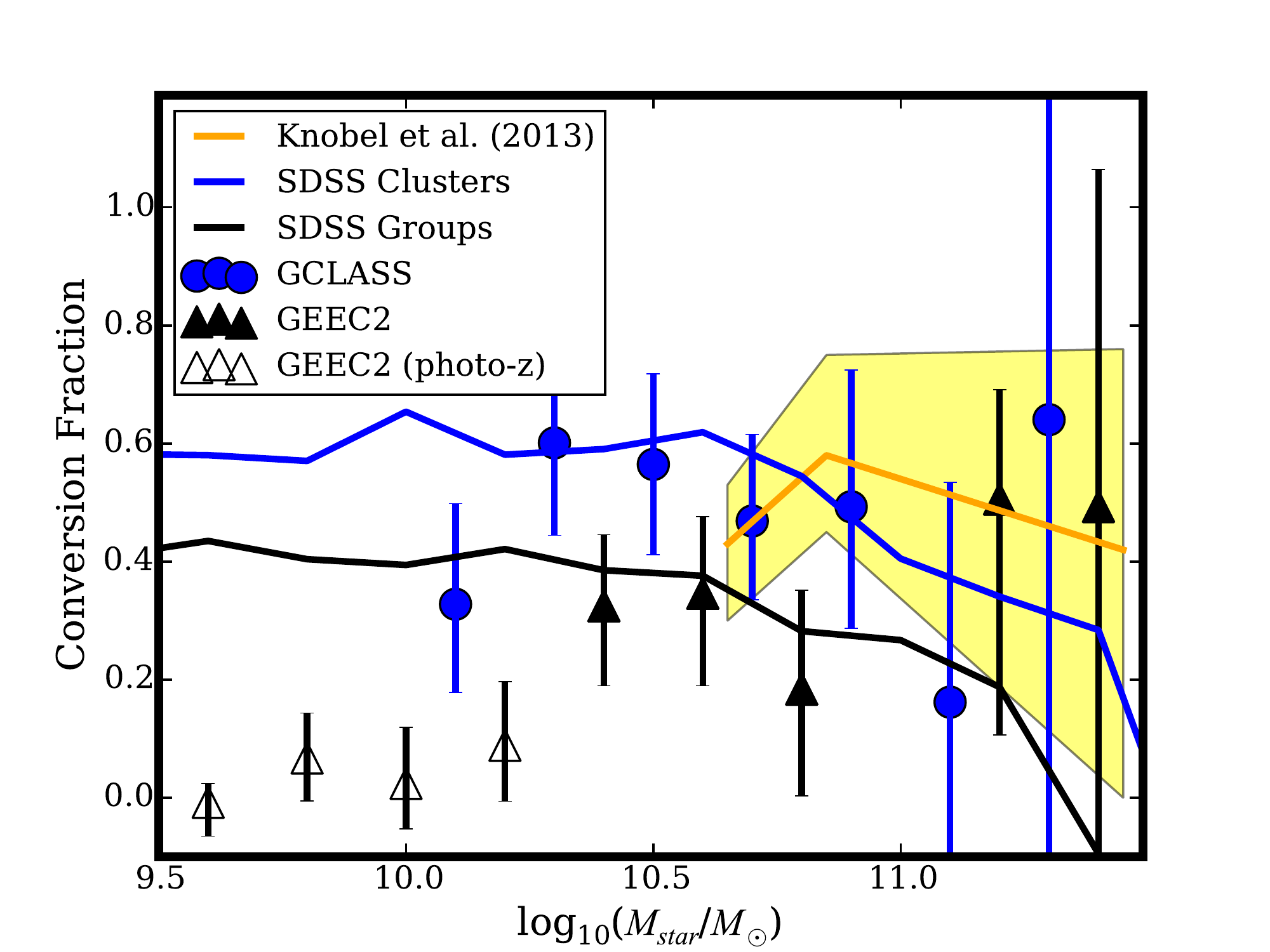}}
        \caption{The environmental conversion fraction (Equation~\ref{eq-qe}) is shown as a function of stellar mass, for GEEC2 and GCLASS.  These are compared with the result of \citet{Knobel_qf}, based on zCOSMOS groups at $z\sim 1$, as the yellow shaded region, and with low-redshift haloes in SDSS from the compilation of \citet{Omand}.  The low redshift samples are chosen to span the same halo mass range (not evolved) as the GEEC2 and GCLASS samples. }
\label{fig:qe}
\end{figure}

Figure~\ref{fig:qe} shows this conversion fraction for GEEC2 and GCLASS, where the field reference sample is again taken from Ultravista at $z=0.9$.  For $M_{\rm star}>10^{10.3}M_\odot$ this quantity is approximately independent of stellar mass: $\sim 60$ per cent in the clusters and $\sim 40$ per cent in the groups.  This is consistent with what others have seen at this redshift \citep[e.g.][]{DEEP2_BO,Knobel_qf,RvdB1}, and we show the \citet{Knobel_qf} results as the yellow shaded region for comparison.  However, when we consider the full stellar mass range accessible to us, including the photometric redshift extension down to $M_{\rm star}=10^{9.5}M_\odot$, we find evidence that \fc\ decreases with decreasing stellar mass, becoming consistent with zero at the lowest masses.  

For comparison, we show the same quantity computed at low redshift, from the SDSS compilation of \citet{Omand}.  For the ``field'' population in Equation~\ref{eq-qe} we use the galaxies classified as ``central'' to their halo.  In Figure~\ref{fig:qe} the solid lines show the value of \fc\ for haloes with mass $13.5<\log{M_{\rm halo}/M_\odot}<14.0$ and $\log{M_{\rm halo}/M_\odot}>14.2$, corresponding to the same halo mass ranges as GEEC2 and GCLASS, respectively. 

It is remarkable that, for stellar masses $M_{\rm star}>10^{10.3}M_\odot$, our measurements of conversion fraction in $z=0.9$ groups and clusters are in excellent quantitative agreement with similar observations at $z=0$, including the fact that more massive haleos have a higher \fc.  This implies that the quenching of massive satellite galaxies in haloes of a given mass is just as effective at $z=1$ as it is at $z=0$, again consistent with the conclusions of \citet{DEEP2_BO} and \citet{Knobel_qf}.  However, the drop in \fc\ observed below $M=10^{10.3}M_\odot$ observed at $z=1$ is in marked contrast to the constancy observed at $z=0$ over this stellar mass range\footnote{We note, though, that significant and dramatic evolution is seen locally at much lower masses \citep[e.g.][]{W+14,Fillingham}}.  This mass dependence may be an important clue to the nature of the physics driving these galaxy transformations, as we will discuss in \S~\ref{sec-physics}.   

\section{Transformation timescales}\label{sec:discuss}
One interpretation of the fact that \fc$>0$ in clusters and groups is that star formation is prematurely truncated as satellite galaxies are accreted, leading to an environmentally--driven transformation.  It is valuable to have a robust estimate of the time for that transformation to occur, as such a timescale could provide important clues about the underlying physical processes responsible.  However, there has been long-standing tension between the relatively short transition timescales derived from the abundance of post-starburst (PSB) and similar galaxies \citep[e.g.][]{P+99,B+99} that appear to be caught in the act of transforming, and the non-zero fraction of star--forming galaxies in clusters and groups which imply a much longer time between accretion and the end of star formation \citep[e.g.][]{McGee11}.  A related observation is that most star--forming galaxies in clusters have specific star formation rates that are indistinguishable from those of field galaxies \citep[e.g.][]{B+04,Baldry06,Mok1,GCLASS,Panstarrs-lin}.
\citet{Wetzel13} suggest that these observations can be reconciled if the total quenching time is made up of a delay time $t_{\rm delay}$, during which there is little change to the SFR, and a final, much shorter ``fading time'' $t_{\rm fade}$ once the SFR begins to decrease.  The delay time at $z=0$ in particular has a nontrivial dependence on stellar mass, likely providing important clues about the physical processes relevant on different scales \citep[e.g.][]{Fillingham}.
Our aim in the following sections is to use our data to constrain the timescale of these separate phases out to $z\sim 1$.

\subsection{Total quenching timescales $t_p$}\label{sec-totaltime}
\begin{figure}
{\includegraphics[clip=true,trim=0mm 0mm 0mm 0mm,width=3.5in,angle=0]{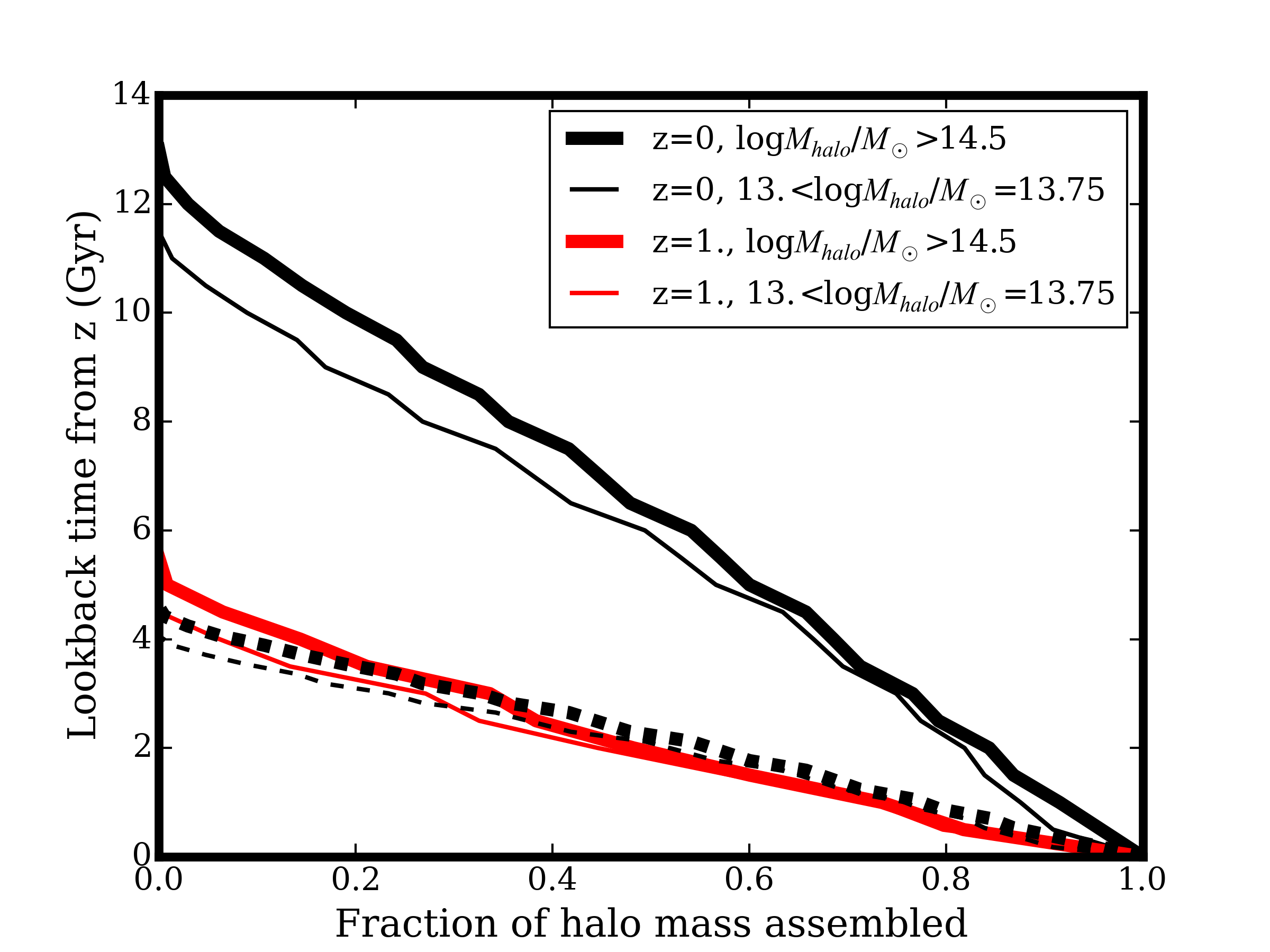}}
        \caption{The time since a galaxy in a halo of a given mass, observed at a given epoch, was first found as a satellite is shown using the formalism of \citet{McGee09}.  The black dotted lines show the $z=0$ curves with the corresponding times rescaled by a factor $(1+z)^{-3/2}$.  This shows that the average accretion rate of haloes of a given mass evolves like the dynamical time.}
\label{fig:McGeeage}
\end{figure}

We follow the work of \citet{McGee09} and \citet{Mok2} to predict the ages of satellite galaxies as a function of halo mass and epoch.  This is based on an analysis of the Millennium simulation \citep{MS}, and the merger trees described by \citet{Helly2003} and \citet{Harker}.  The semi-analytic model of \citet{Bower06}, updated to use the more realistic strangulation prescription of \citet{Font}, is used to track the assembly of stellar mass within these haloes. Galaxies are identified as central or satellite, and we identify the lookback time $t_{\rm assemble}$ at which any observed galaxy first became a satellite, in any halo.   Figure~\ref{fig:McGeeage} shows $t_{\rm assemble}$ as a function of halo mass and redshift.  It shows, for example, that half the cluster members at $z=0$ have been satellites for $\sim 7$ Gyr, while at $z=1.0$, this fraction corresponds to $\sim 2$ Gyr.   There is a relatively small halo mass dependence as well; typically galaxies in cluster-mass haloes were accreted $\sim 10$ per cent longer ago than those in group-mass haloes.  

We will make the assumption that the accretion history of galaxies is independent of their stellar mass or other properties, and we neglect any differential evolution in stellar mass, or tidal disruption of satellites.  In that case we can interpret \fc\ from Figure~\ref{fig:qe} as the fraction of initially star-forming galaxies for which the time between accretion and final quenching ($t_p$) is less than the time between accretion and observation ($t_{\rm assemble}$).  This allows us to associate \fc\ with the x-axis of Figure~\ref{fig:McGeeage}, and thus determine $t_p$.  For example, the lowest-mass bin of the GCLASS data is \fc\ $\sim 0.3$, and this is interpreted to mean that 30\% of accreted, star--forming galaxies had their star formation prematurely truncated through environmental processes.  Figure~\ref{fig:McGeeage} shows that 30\% of galaxies in $z=1$ clusters have been satellites for at least $\sim 3$ Gyr.  Thus we conclude that $t_p$=3 Gyr, in this example.

\begin{figure}
{\includegraphics[clip=true,trim=0mm 0mm 0mm 0mm,width=3.5in,angle=0]{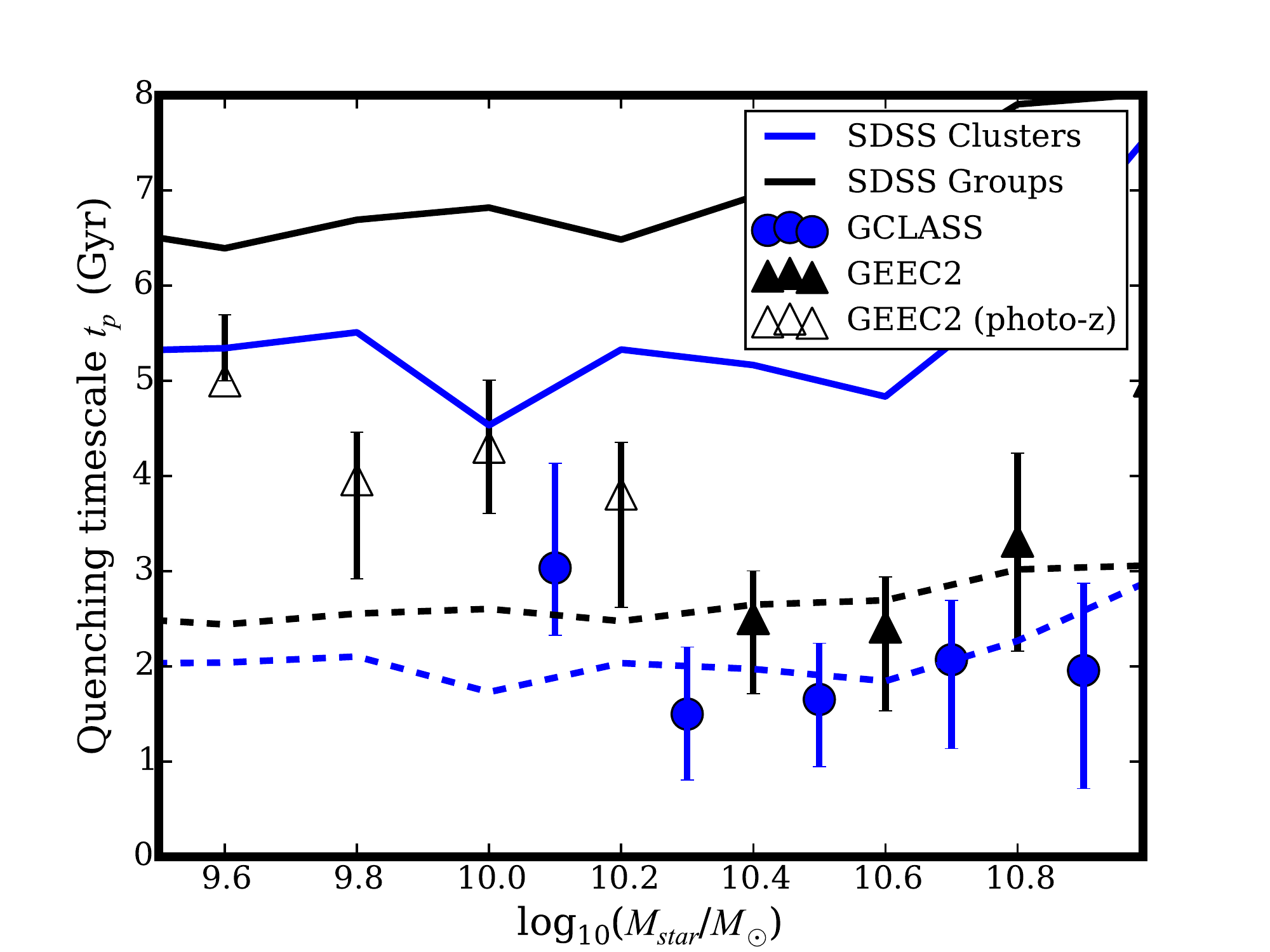}}
        \caption{Using the accretion histories shown in Figure~\ref{fig:McGeeage} we convert the observed \fc\ into the time $t_p$, which represents the time between a galaxy first becoming a satellite and when it is finally classified as passively evolving.  The solid lines show the $z=0$ results from SDSS, and the dashed lines show the same curves rescaled by $(1+0.9)^{-1.5}$.  These latter are the predicted relations at $z=0.9$ if the timescale evolves like the dynamical time.}
\label{fig:tptf0}
\end{figure}

With this transformation from \fc\ to $t_p$, we present Figure~\ref{fig:tptf0}.  Since \fc\ is poorly constrained at $M>10^{11}M_\odot$, we will restrict this and all further analysis to lower mass galaxies.  Locally, $t_p\gtrsim 5$ Gyr, approximately independent of stellar mass but with a weak dependence on halo mass, such that it increases to $\sim 6.5$ Gyr for group--scale haloes.  These are shown as the blue and black solid lines, based on our SDSS analysis.
Our new measurements at $z=0.9$ from GCLASS and GEEC2 are shown as the circles and triangles, respectively\footnote{Note that the resulting timescales are all longer than the corresponding estimates given by \citet{Wetzel13} and \citet{Mok2}.  The difference can be attributed to different assumptions about the population composition at the time of accretion, and we discuss this in Appendix~\ref{sec-app2}.}.   The observed trend of \fc\ with stellar mass translates into a strong trend in $t_p$, increasing from $\sim 2$ Gyr to $\sim 5$ Gyr with decreasing stellar mass. 

For massive galaxies, $M_{\rm star}>10^{10.3}M_\odot$, the lack of evolution in \fc\ noted in Figure~\ref{fig:qe} in fact demands a strong evolution in $t_p$, such that $t_p$ decreases with redshift.  This is because the higher redshift satellite population must have been assembled over a much shorter time.  The magnitude of this evolution is in approximate agreement with the evolution in dynamical time, as previously noted by others \citep[e.g.][]{TW10,Mok2}.  We show this explicitly by scaling the low redshift measurements of $t_p$ by $(1+z)^{-3/2}$, and replotting as the dashed lines in Figure~\ref{fig:tptf0}.  

However, this dynamical time scaling does not appear to apply to lower-mass galaxies, for which we observe a steadily increasing value of $t_p$ with decreasing stellar mass, reaching a value of $\sim 5$ Gyr that is comparable to the measurements in groups at $z=0$, and approaches the Hubble time at $z=0.9$, $t=6.3$ Gyr. 
Moreover, if the model of \citet{Wetzel13} is correct, then $t_p=t_{\rm delay}+t_{\rm fade}$, and the delay and fading timescales are likely driven by different physical processes.  The fact that $t_p$ scales roughly with dynamical time for massive galaxies does not imply that the same is necessarily true for $t_{\rm delay}$ or $t_{\rm fade}$.  We discuss this further in \S~\ref{sec-physics}.

\subsubsection{Transition galaxies and fading timescales}\label{sec-transition}
If star formation in satellite galaxies is prematurely quenched relative to central galaxies, there should exist an excess population of galaxies with lower-than-average star formation rates, and their abundance can be related to the timescale for this star formation to shut down.  For the $z=0$ population, \citet{Wetzel13} measured an exponential fading timescale from the abundance of galaxies with low but non-zero SFR.  They found that this timescale depends on stellar mass, ranging from $\tau_q=0.3\pm 0.2 \mbox{ Gyr}$ at $M_{\rm star}=5\times 10^{10}M_\odot$ to $\tau_q=0.6\pm 0.2$ Gyr at $M_{\rm star}=2\times 10^{10}M_\odot$.  For the purposes of comparing with our data, where we have insufficient statistics to consider trends with stellar mass, we will adopt $\tau_q=0.5\pm 0.4$ Gyr.  Assuming that a drop in SFR by a factor $3$ would be sufficient for a galaxy to be classified as passive, $t_{\rm fade}\sim \tau_q$.

In both GCLASS \citep{GCLASS} and GEEC2 \citep{GEEC2,Mok1,Mok2} we attempted to identify similar populations of galaxies intermediate between the normal star--forming and passively-evolving populations.  While both studies came to similar conclusions, the two populations were identified in very different ways.  Here we present a self-consistent analysis using the \citet{GCLASS} definition of ``transition'' galaxies with declining SFR (which they call ``poststarburst''), since it is possible to apply this to both samples in a similar way.  These are identified as blue galaxies, with $\mbox{D4000n}<1.45$, but without detectable [OII] emission lines\footnote{The D4000n index, as defined by \citet{B+99}, is insensitive to dust extinction or metallicity effects.  Given the deep spectroscopy, and relatively shallow imaging, available for GCLASS, this feature was deemed the best way to identify galaxies with young populations.}.     To identify galaxies without significant [OII] emission in GEEC2 we make a selection on rest-frame equivalent width, $W_\circ$(OII)$<3$\AA, which corresponds approximately to the detection limit in GCLASS.

From the stacked spectra of transition galaxies, \citet{Muzzin-ps} concluded that they are consistent with a SFR that declines linearly to zero over a timescale $0.4^{+0.3}_{-0.4}$ Gyr, very similar to $t_{\rm fade}$ measured at $z=0$. A complementary approach is to attempt to determine the total amount of time galaxies spend in the poststarburst phase.  This can be achieved by comparing their abundance with the number of star--forming galaxies at the same mass.  
Note that the total quenching time presented in Figure~\ref{fig:tptf0}, $t_p$, also represents the time during which all presently star--forming and transition satellite galaxies were accreted.   We then assume that the fraction of these blue galaxies that are in the transition phase is equal to the fraction of $t_p$ spent in that phase.  That is:
\begin{equation}\label{eqn-psbtime}
\frac{t_{\rm fade}}{t_{\rm SF+trans}}=\frac{t_{\rm fade}}{t_p}=\frac{N_{\rm trans}}{N_{\rm SF}+N_{\rm trans}},
\end{equation}
where the {\it trans} subscripts refer to galaxies in the transition population.

\begin{table}
\begin{tabular}{llllll}
\hline
\hline
Sample &SF& trans & $\frac{N_{\rm trans}}{N_{\rm SF+trans}}$&$t_p$&$t_{\rm fade}$\\
       &  &     &                                     &(Gyr)& (Gyr)\\
\hline
GEEC2  &12& 7   & $0.37\pm0.1$&$2.4\pm0.6$&$0.9\pm0.3$\\
GCLASS &18& 8   & $0.31\pm0.09$&$1.7\pm0.6$&$0.5\pm0.2$\\
\hline
\end{tabular}
\caption{This table summarizes the abundance of star--forming and transition galaxies with $r<R_{200}$ and $10^{10.3}<M_{\rm star}/M_\odot<10^{10.75}$, for each sample.  Column (4) gives the fraction of recently accreted galaxies that are in the transition phase, from which we can estimate the timescale for SFR to decline to zero ($t_{\rm fade}$).\label{tab-abundances}}
\end{table}

In Table~\ref{tab-abundances} we show the number of star--forming and transition galaxies in each sample, with $r<R_{200}$ and limited to the stellar mass range $10^{10.3}<M_{\rm star}/M_\odot<10^{10.75}$, since the GEEC2 target selection is biased against lower-mass galaxies with colours typical of this population.  These transition galaxies make up a relatively large fraction of the blue population in both samples, $\sim 0.35\pm0.1$.  We read $t_p$ in the relevant mass range from Figure~\ref{fig:tptf0}, and therefore find $t_{fade}=0.9\pm0.3$ Gyr in GEEC2 and $t_{fade}=0.5\pm 0.2$ Gyr in GCLASS.  Though the statistical and systematic uncertainties associated with these estimates are large, it is notable that the two timescales are consistent with each other, with the independent analysis of spectral features in \citet{Muzzin-ps}, and with the corresponding measurement at $z=0$.  However, we cannot rule out the possibility of $t_{\rm fade}\sim 0.2$Gyr, which would be expected if it evolved like the dynamical time.

\subsubsection{Delay times}\label{sec-tdelay}
\begin{figure}
{\includegraphics[clip=true,trim=0mm 0mm 0mm 0mm,width=3.5in,angle=0]{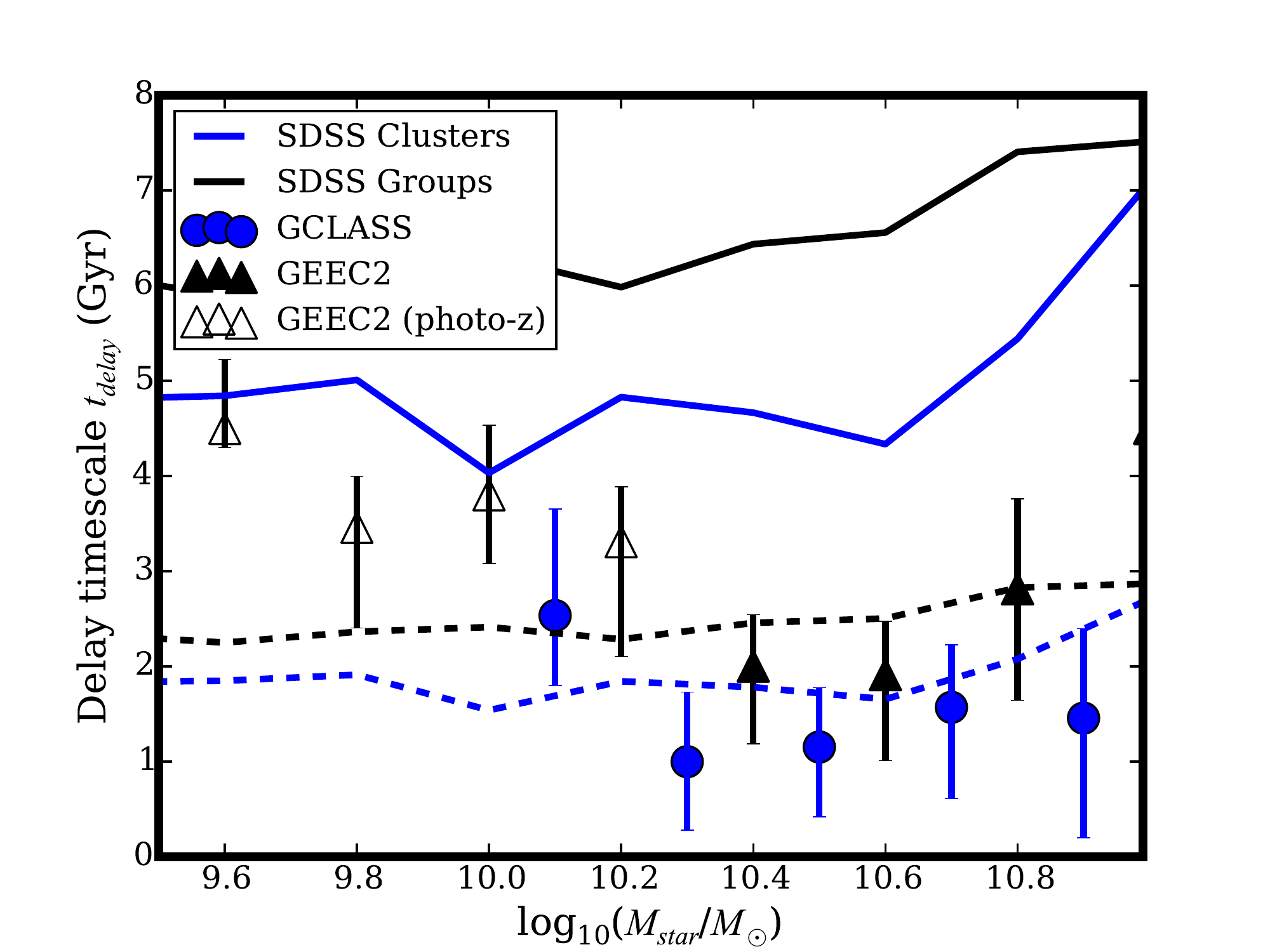}}
        \caption{Derived from the values of $t_p$ presented in Figure~\ref{fig:tptf0}, and assuming a constant $t_{\rm fade}=0.5\pm 0.2$ Gyr, we show the delay time $t_{\rm delay}$ as a function of redshift and halo mass.  Again the solid lines show the $z=0$ results from SDSS, and the dashed lines show the same curves rescaled by $(1+0.9)^{-1.5}$, corresponding to dynamical time evolution.  }
\label{fig:tptf0.5}
\end{figure}

Similar to what is observed locally, the fading times we derive at $z\sim 0.9$ are considerably shorter than the total estimated time between accretion and the final cessation of star formation, $t_p$.  The implication is that there is a delay time $t_{\rm delay}$, during which star formation continues unabated.  In Figure~\ref{fig:tptf0.5} we show the derived $t_{\rm delay}$, assuming a redshift- and mass-independent $t_{\rm fade}=0.5\pm0.2$ Gyr.  This is equivalent to Figure~\ref{fig:tptf0}, with all times reduced by $t_{\rm fade}$ and the additional uncertainty added in quadrature.

\section{Discussion}\label{sec-physics}
The most common interpretation of environmentally-induced quenching of star formation is that it is due to the removal of gas (hot or cold) through various processes related to the interaction between satellite galaxies and the massive halo.  There are numerous viable gas-removal mechanisms operating in dense environments, including ram pressure, tidal forces, and evaporation.  In all these cases, the simplest expectation is that any delay timescale is related to the galaxy's orbit through the halo potential, and should scale with redshift in a way that traces the evolution in halo dynamical time.

However, \citet{overconsumption} pointed out that, especially at higher redshift, the gas consumption timescale may be dominated by the balance between cosmological accretion rates and gas ejection processes.  In particular, the high star formation rates, which in most models require similarly high gas ejection rates, result in short gas depletion timescales in the absence of cosmological accretion.  Because the average star formation rate depends on redshift and galaxy mass in a way that is largely decoupled from the properties of the host halo \citep[e.g.][]{Behroozi13}, this results in depletion timescales that are not simply related to the halo dynamical time.  

In the equilibrium model of \citet{overconsumption}, following the spirit of \citet{Dave12} and \citet{Lilly13}, the inflow of gas onto a galaxy is balanced by the outflow rate and the rate of star formation. In the absence of cosmological accretion, appropriate for satellite galaxies, we associate the delay time with the time required to deplete a galaxy's gas ``reservoir'', $t_{\rm delay}=M_{\rm res}/{{\dot M}_{\rm res}}$.  Here the reservoir includes any gas associated with the galaxy upon infall that is potential fuel for future molecular cloud formation.    In the expansion below we evaluate both $M_{\rm res}$ and ${\dot M}_{\rm res}$ at the epoch of observation, $z$.  At the time of infall $z_{\rm infall}$, both quantities are expected to be larger, and a more accurate calculation of $t_{\rm delay}$ requires integrating the solution between $z_{\rm infall}$ and $z$.  We neglect this in order to retain the transparent simplicity of the model; it has a small quantitative effect on our results but does not change our conclusions.
Finally, the identification of this timescale as $t_{\rm delay}$ requires that the SFR does not significantly change as the reservoir is depleted; if this is true it has important implications for the nature of the reservoir, which we otherwise disregard here.

Making the assumption that the amount of gas permanently ejected from the halo is proportional to the star formation rate, ${\dot M}_{\rm ej}=\eta {\dot M}_{\rm star}$, leads to an expression for the delay time given by \citet{overconsumption}:
\begin{equation}\label{eqn-oc}
t_{\rm delay}=\frac{f_{\rm baryon}-f_{\rm cold}-f_{\rm star}\left(1+\eta\left(1+R\right)\right)-f_{\rm strip}}{f_{\rm star}\left(1-R+\eta\right)sSFR},
\end{equation}
where $f_{\rm baryon}=0.17$, $f_{\rm cold}$ and $f_{\rm star}$ are the fraction of the halo mass found in baryons, cold (molecular) gas and stars, respectively.  $f_{\rm strip}$ is the mass in baryons, expressed as a fraction of the halo mass, that might have been stripped from the satellite galaxy (by ram pressure, for example); unless otherwise specified we set $f_{\rm strip}=0$ to estimate the upper limit on delay time, in the absence of any dynamical effects.  Finally $R$ is the fraction of gas that is instantaneously recycled into the ISM upon star formation, and sSFR is the specific star formation rate.  Following \citet{overconsumption}, we use empirical determinations of $f_{\rm star}(M_{\rm halo})$ and $sSFR(M_{\rm star},z)$ from \citet{Behroozi13}, and $f_{\rm cold}=0.1\left(1+z\right)^2f_{\rm star}$ from \citet{CW13}.  Adopting a Chabrier IMF with an appropriate $R=0.4$ \citep{overconsumption} leaves us with one free parameter, $\eta$, which for simplicity we assume to be independent of halo mass and redshift.  

\begin{figure}
{\includegraphics[width=3.5in,angle=0]{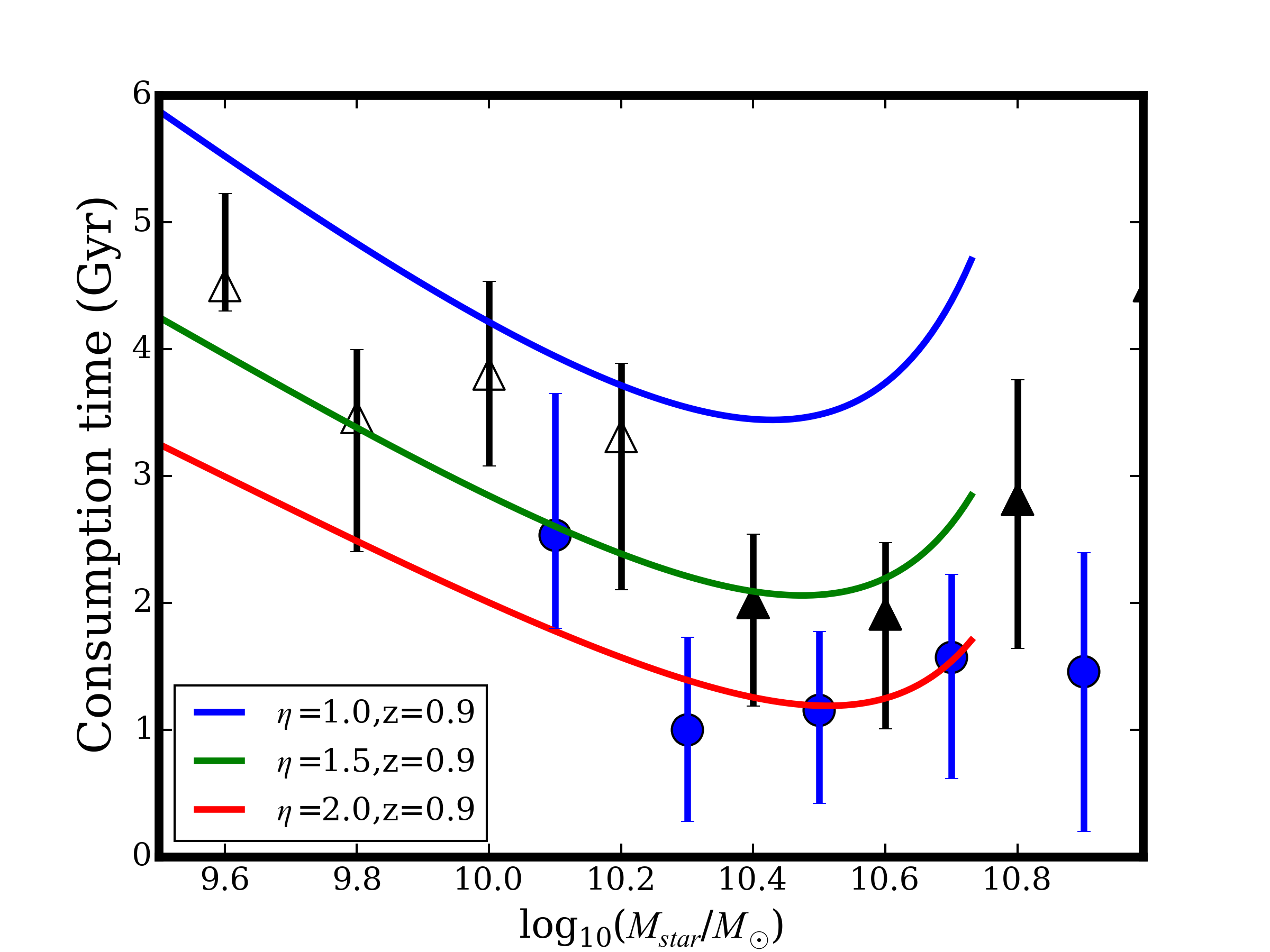}}
        \caption{Lines show the maximum time over which a satellite galaxy of a given stellar mass, at redshift $z=0.9$, could maintain its SFR unchanged once cosmological accretion of matter has stopped.  The time is measured assuming that any depletion of the reservoir does not reduce the SFR, following \citet{overconsumption}.   These models make the simple assumption that star formation is accompanied by outflows that permanently carry away mass at a rate $\eta$SFR from the galaxy, with $\eta$ independent of mass and redshift. We compare with our GCLASS (blue circles) and GEEC2 (triangles) data over the relevant stellar mass range, and find the observed stellar mass dependence is in remarkably good agreement with the $1.5<\eta<2.0$ model predictions.  The small halo-mass dependence might indicate that dynamical effects still play a secondary role in the more massive clusters.}
\label{fig:oc}
\end{figure}

For a given value of $\eta$, this model predicts that $t_{\rm delay}$ should depend strongly on stellar mass and redshift,
 in a way that is qualitatively different from naive expectations due to any dynamical stripping processes.  In this way, despite its simplicity, it provides a useful alternative with which to compare observations.    In Figure~\ref{fig:oc}
we show this prediction at $z=0.9$, as a function of stellar mass for different values of $\eta$.  These predictions\footnote{We compute the model only for halo masses $M<2\times10^{12}M_\odot$.  At higher masses, most of the reservoir is in a very hot phase (e.g. the intracluster medium) where the cooling time exceeds a Hubble time; the simple model does not account for that and leads to a large overestimate of $t_{\rm delay}$.}
 are compared directly with our measurements of delay time from Figure~\ref{fig:tptf0.5}. 

  We find that the overconsumption model provides a remarkably good description of the data, including the strong stellar mass dependence.  For the groups the data are best fit with $\eta\sim 1.5$, while the higher values of \fc\ observed in the more massive clusters prefer $\eta\sim 2.0$.  In either case, as noted by \citet{overconsumption}, this is a rather small factor, implying that only a modest amount of gas is permanently ejected from galaxies.    Note that this comparison assumes that cosmological accretion stops once a galaxy becomes a satellite of a larger halo.  In reality accretion of dark matter may stop earlier than this \citep[e.g.][]{Behroozi14}; in that case the consumption times indicated by our data should be longer than shown, implying $\eta$ is even lower.   Similarly, adopting a more sophisticated approach where we integrate $M_{\rm res}/{{\dot M}_{\rm res}}$ over the time between infall and observation also leads to smaller values of $\eta$.  On the other hand, some hydrodynamic simulations have shown that at least massive satellites might continue to accrete additional gas after becoming a satellite, which would lead to an underestimate of $\eta$ \citep[e.g.][]{Keres}.  
Importantly, in this model $\eta$ is representative of permanent gas expulsion rates from all galaxies, not only satellites. In this way we are able to use the differential comparison of satellite and central galaxies to put new constraints on how star formation and feedback operates in central galaxies. 

The good qualitative agreement between the data and the models with $\eta\sim 1.5$ permits an interesting hypothesis about the origin of the delay time and ultimate quenching of star formation in satellite galaxies at $z\sim 1$.  In particular, that dynamical stripping processes are not required to explain the observations.  Instead, simply shutting off the inflow of new gas through cosmological accretion limits the star--forming lifetime of a galaxy, as it consumes its reservoir through star formation and permanently expels gas with a relatively modest factor of $\eta$. The high star formation rates of galaxies with $M_{\rm star}\sim10^{10.5}M_\odot$ leads to particularly short delay times.  

This does not mean dynamical processes play no role at all, and analysis of the velocity distribution of post-starburst galaxies in GCLASS in fact provides some direct evidence for such processes in the more massive clusters \citep{GCLASS-Noble,Muzzin-ps}.  If dynamical processes act to remove some gas, then in our model $f_{\rm strip}>0$ and a lower $\eta$ (potentially zero) is required to fit the data.  Assuming a modest $f_{\rm strip}=0.03$ (corresponding to 18 per cent of the baryonic mass) in the highest mass haloes allows them to be modeled with the same $\eta=1.5$ that works for the groups.  The consequence of assuming a significant $f_{\rm strip}$ in all haloes, though, would be that outflows must eject a surprisingly small amount of mass.  For example, if at least half the baryonic mass was stripped ($f_{\rm strip}<0.085$), then we would require $\eta<0.5$ to match the data.  These theoretical uncertainties and simplifications, together with remaining possible systematic effects in the data itself (e.g. possible incompleteness at low stellar masses) and derived quantities (e.g. neglecting stellar mass growth after becoming a satellite), mean that the absolute value of $\eta$ found here should be considered indicative, rather than definitive.  With larger samples and deeper data a more sophisticated analysis will be warranted.

At $z=0$, the low star formation rates, and associated gas ejection rates, mean the gas consumption timescale\footnote{Recall that this is the time required to consume all gas available to the galaxy, including its reservoir of warm gas.  Thus it is much longer than the cold gas consumption timescale.} computed from Equation~\ref{eqn-oc} is much longer, $>10$ Gyr at $M_{\rm star}=10^{10.5}$ \citep{overconsumption}.  This is well above the measured $t_{\rm delay}\sim 5$ Gyr for comparable values of $\eta$, and also much longer than the dynamical time.  Overconsumption is therefore unlikely to be important today, and the shutdown of star formation in recently accreted satellites is most likely driven by the well-studied dynamical processes like ram pressure stripping.  This leads to a fundamentally different dependence of $t_{\rm delay}$ on galaxy stellar mass. It also affords a natural explanation for the observed halo-mass dependence at $z=0$, if such dynamical effects are more prevalent in higher-mass haloes.

\section{Conclusions}
We have presented a joint spectroscopic analysis of 20 massive galaxy haloes at $0.8<z<1.2$, selected from the GCLASS \citep{GCLASS} and GEEC2 \citep{geec2_data} surveys.  These are compared with the field at the same redshift from Ultravista \citep{Ultravista}, and with observations at $z=0$ compiled from SDSS data by \citet{Omand}.  Our main conclusions can be summarized as follows.
\begin{itemize}
\item Satellite galaxies in clusters at $z=0.9$ have a higher fraction of passive galaxies than galaxies of similar stellar mass in the field, and comparable to the fractions seen locally.  The GEEC2 groups show a more modest difference compared to the surrounding field, and the difference disappears entirely for stellar masses below $M_{\rm star}=10^{10.3}M_\odot$.
\item We define \fc\ to represent the fraction of star-forming satellites that stopped forming stars prematurely for their stellar mass, and find \fc$\sim 0.4$ in the groups and $\sim 0.6$ in the clusters, for $M_{\rm star}>10^{10.3}M_\odot$. We find evidence that \fc\ may increase with increasing stellar mass, in contrast with what is observed at $z=0$ for this stellar mass range. 
\item From the abundance of blue galaxies without [OII] emission we estimate the fading time, during which galaxies in the blue cloud have reduced star formation rates, and find that it is consistent with $\sim 0.5\pm 0.2$ Gyr in both groups and clusters at $z=0.9$.  This is comparable to the fading time estimated at $z=0$.
\item To reconcile the observed \fc\ with the short fading timescale requires a delay time after accretion that depends on stellar mass.  For massive galaxies, $M_{\rm star}>10^{10.3}M_\odot$, $t_{\rm delay}$ is substantially shorter than the equivalent times at $z=0$.  However, $t_{\rm delay}$ increases with decreasing stellar mass, approaching the $z=0$ value of $\sim 5$Gyr at $M_{\rm star}\sim 10^{9.5}M_\odot$.  
\item We compare these results with the simple analytic model of \citet{overconsumption}, where the evolution of central galaxies is determined empirically with a single free parameter $\eta$, which is the ratio of permanently ejected mass to star formation rate.  Satellite galaxies are assumed to obey the same physics, but without a source of cosmological accretion.  We find our data are in good agreement with the predictions of this model, with $\eta\sim 1.5\pm 0.5$.   
\end{itemize}

The observations suggest that the mechanisms for quenching satellite galaxies may be fundamentally different at $z=0.9$ and $z=0$.  At the higher redshift, the lack of cosmological accretion, combined with high rates of star formation and modest mass ejection, lead to exhaustion of fuel on a short timescale that depends strongly on galaxy mass.  This same process at $z=0$ is much less efficient, and cannot easily explain the observed properties of nearby satellite galaxies for which \fc\ is independent of stellar mass. Instead, it is likely that dynamical processes like ram pressure stripping become important.

There is considerable value in extending this analysis to higher redshift.  If satellite quenching depends on both orbital characteristics and internal feedback/outflow rates \citep{overconsumption}, we can use the fact that the associated timescales evolve in very different ways.  Galaxies at $z>1$ have even higher star formation and outflow rates, and it may in fact be expected that satellites beyond some redshift are all quenched very quickly as a result.

Although we have made an attempt to interpret the quenching timescales measured from different surveys self-consistently, there are many complicating factors.  We have no explicitly addressed, for example, the question of when cosmological infall ceases, which defines the start of any delay time.  We have also largely ignored the role of tidal disruption and merging, and stellar mass growth after accretion.  The estimate of $t_{\rm fade}$ is statistically and systematically uncertain, and its dependence on stellar mass is unconstrained at $z>0$.  Finally the sample of groups and clusters at $z\sim 1$ is still small and their selection may be prone to a progenitor bias, such that they are the more evolved systems for their masses.  We have shown that careful study of satellite galaxies can reveal insight not only into the causes of premature quenching, but into the more fundamental interplay between gas accretion and outflows in {\it all} galaxies.   The potential power inherent in measuring the evolution in these timescales means future work aimed at further understanding and reducing these systematic effects will be worthwhile.

\section{Acknowledgments}
\par
MLB would like to acknowledge generous support from NOVA and NWO visitor grants, as well as the excellent hospitality extended by the Sterrewacht of Leiden University, where this paper was mostly written during sabbatical leave from Waterloo.
We are grateful to the SDSS, COSMOS and zCOSMOS teams for making their excellent data products publicly available. This research is supported by NSERC Discovery grants to MLB and LCP.  RFJvdB acknowledges support from the Netherlands Organisation for Scientific Research grant number 639.042.814, and the European Research Council FP7 grant number 340519.  DJW acknowledges the support of the Deutsche Forschungsgemeinschaft via Project ID 3871/1-1.  AF wishes to acknowledge Finnish Academy award, decision 266918.
\bibliography{ms}
\appendix
\section{Upper limits on passive galaxy content}\label{sec-app1}
Below the spectroscopic limit of GEEC2, we rely on photometric redshifts to estimate membership.  Rather than relying on statistical background subtraction, we integrate the $p(z)$ probability distribution function to assign a probability that a galaxy is in a group, as described in \citet{geec2_data}.  There are several potential problems with this approach.  One of course is that it assumes the $p(z)$ is correct; furthermore we are only approximating the distribution as two semi-Gaussian distributions, based on the 68$^{th}$ percentile uncertainties.  Probably more significant for us is the fact that the actual $p(z)$ will be modified by the prior knowledge that there is an overdensity of galaxies in the field.  In \citet{geec2_data} we attempt to make a global correction for this, based on the spectroscopic sample.  It is possible, and even likely, that the correction should depend on galaxy properties like stellar mass, apparent magnitude, and colour.  

Since several of our results are driven by the low fraction of passive group members below the spectroscopic limit, here we wish to consider an upper limit on that fraction, by adopting a conservative treatment of the photometric members.  Specifically, we will consider all passive galaxies that have $p_g>0.1$ to be group members.  Here, $p_g$ is already corrected (i.e. larger) for the overdensity bias noted above.  In other words, we assume that every passive galaxy that has a 10\% chance of being in the group is actually in the group; this is surely an upper limit.  To get an upper limit on the passive fraction (and hence \fc) we must determine a lower limit on the number of star--forming members.  To do this we will take the probability from the integrated $p(z)$, without correcting for overdensity bias.  

\begin{figure}
{\includegraphics[clip=true,trim=0mm 0mm 0mm 0mm,width=3.5in,angle=0]{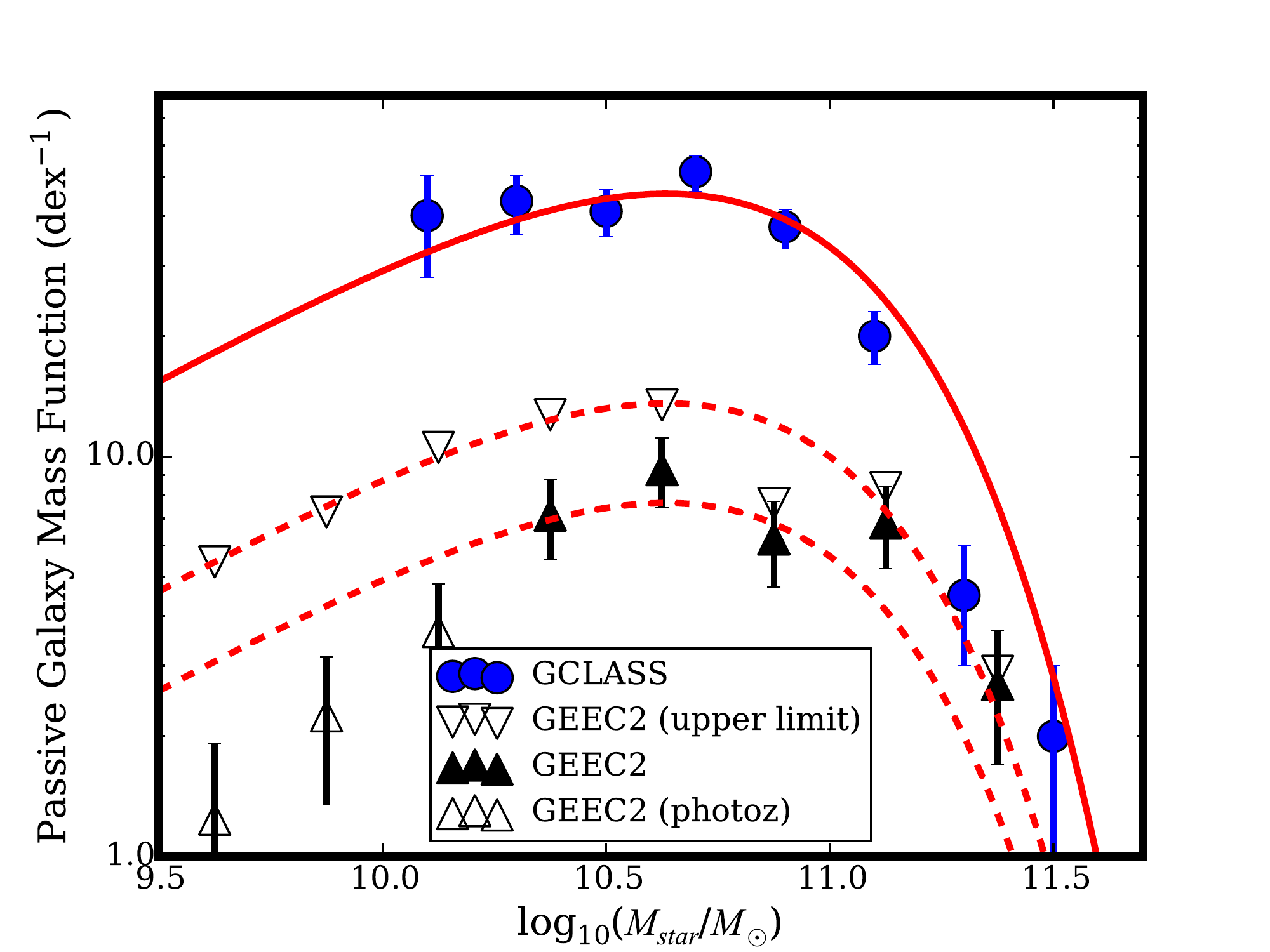}}
        \caption{This figure is the same as Figure~\ref{fig:pmf}, but including conservative upper limits on the group galaxy passive mass function, as inverted open triangles.  The Schechter function representing the same population in the field is shown renormalized both to the measured data (filled and open triangles), and to our upper limits.}
\label{fig:pmf_ulim}
\end{figure}

In Figure~\ref{fig:pmf_ulim} we show the passive galaxy mass function from Figure~\ref{fig:pmf}, and compare it with the upper limits (inverted triangles) computed as described above.  The shape defined by the upper limits is now in better agreement with the renormalized field galaxy mass function shape (upper dashed line), with a faint end slope $\alpha\sim -0.4$.  Note that, at high stellar masses where the spectroscopic completeness is high, the upper limits are not far above the measurements (filled triangles).  

\begin{figure}
{\includegraphics[clip=true,trim=0mm 0mm 0mm 0mm,width=3.5in,angle=0]{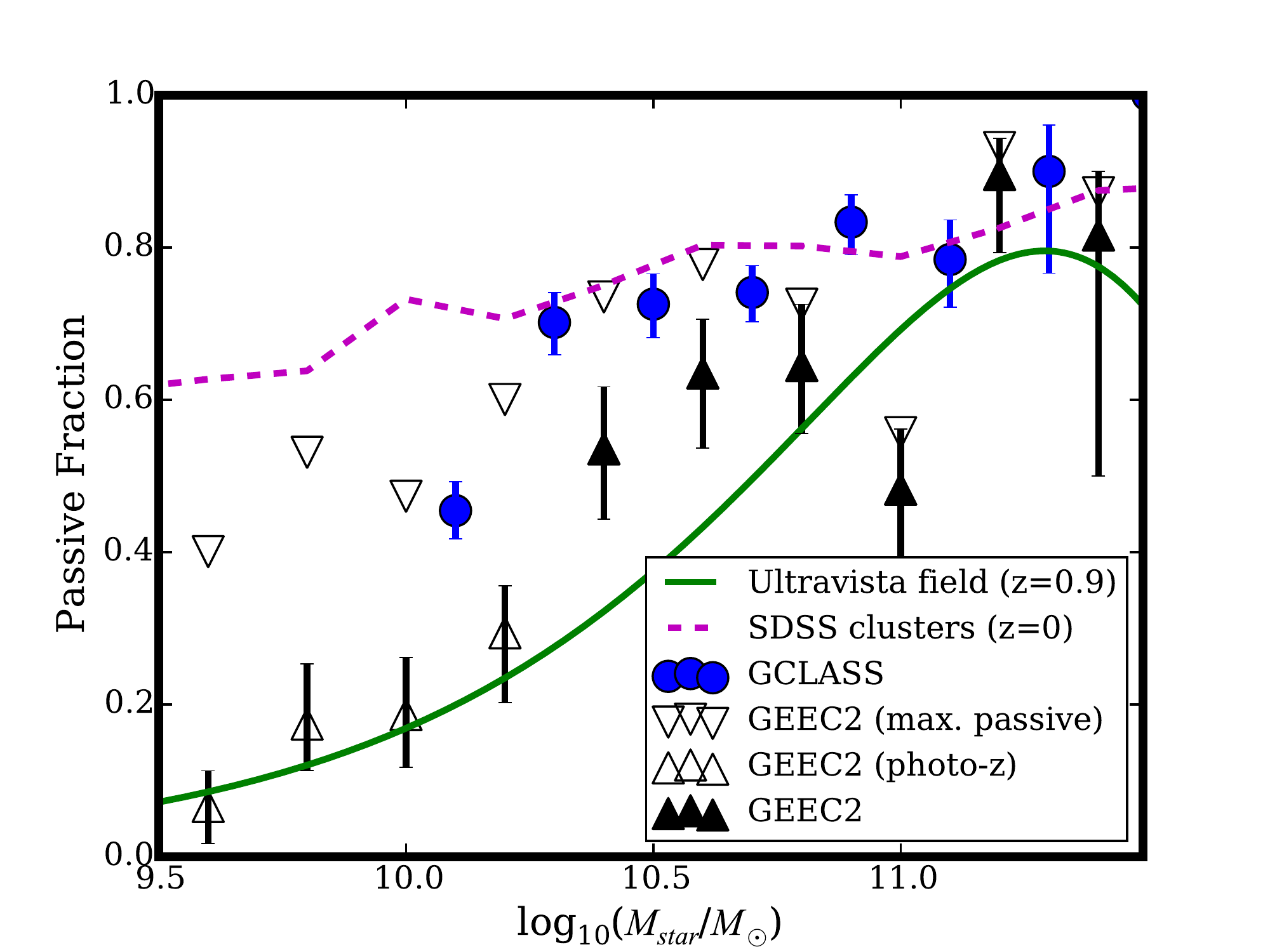}}
        \caption{As Figure~\ref{fig:qf}, but including the upper limits on the GEEC2 quenched fraction as inverted triangles. These upper limits are consistent with the GCLASS measurements and, at low stellar masses, lie between the local values and our best estimates (triangles).}
\label{fig:qf_ulim}
\end{figure}

Figure~\ref{fig:qf_ulim} shows the quenched fraction in GCLASS and GEEC2, as in Figure~\ref{fig:qf}, now including the conservative upper limits on the GEEC2 photometric redshift sample as inverted triangles.  These upper limits are in good agreement with the GCLASS data.  While they still deviate from the $z=0$ results at low stellar masses, the fractions are considerably higher than inferred in our best estimate.  

\begin{figure}
{\includegraphics[clip=true,trim=0mm 0mm 0mm 0mm,width=3.5in,angle=0]{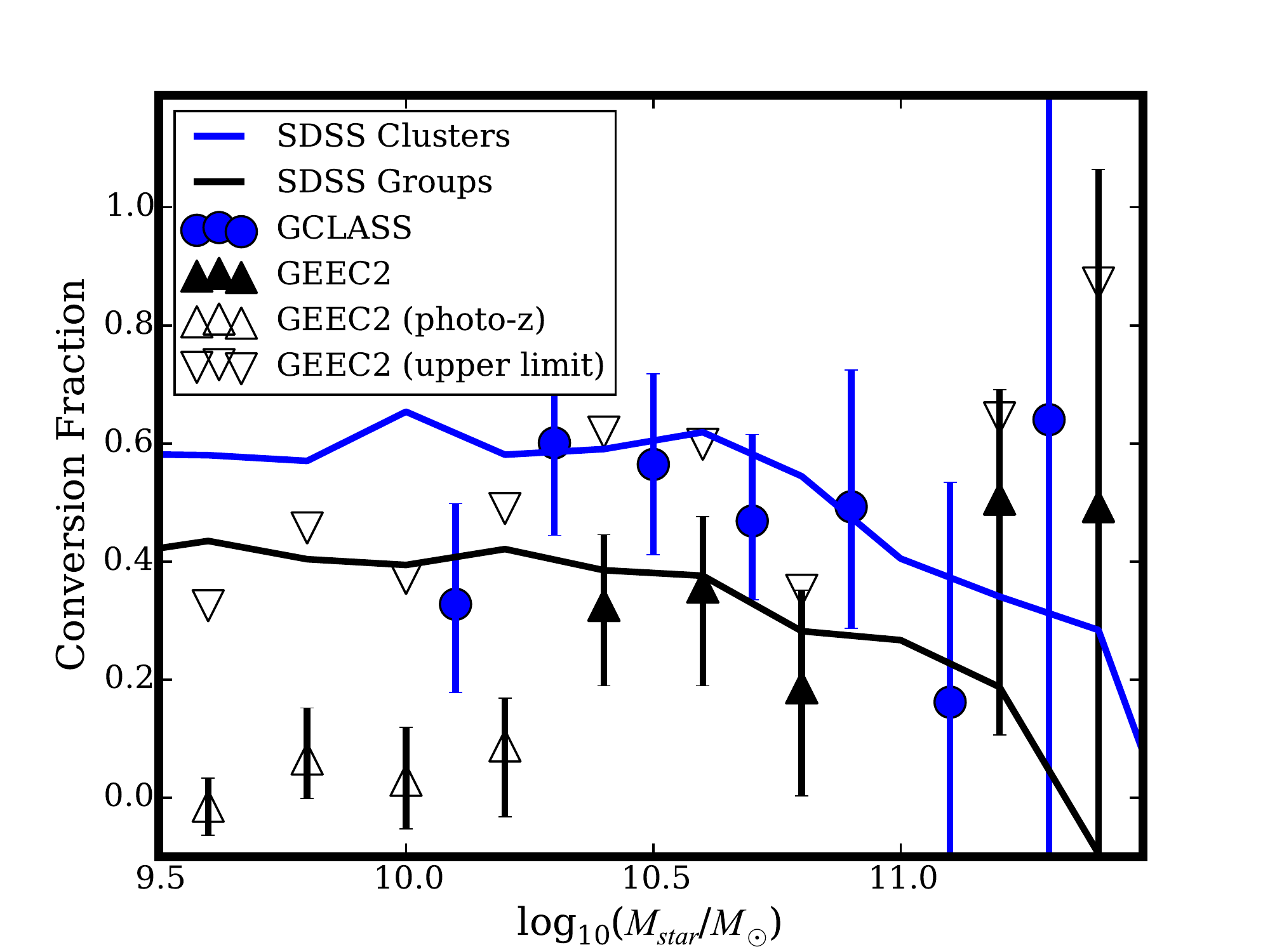}}
        \caption{As Figure~\ref{fig:qe}, the environmental conversion fraction (Equation~\ref{eq-qe}) is shown as a function of stellar mass, here including conservative upper limits on the GEEC2 measurements (inverted triangles).  }
\label{fig:qe_ulim}
\end{figure}

Finally, in Figure~\ref{fig:qe_ulim}, we reproduce Figure~\ref{fig:qe}, again including the upper limits for GEEC2.  As expected from the previous Figure, the limits are consistent with the GCLASS measurements.  At lower masses they remain consistent with the corresponding $z=0$ measurements of similar-mass groups from SDSS.

Even in this conservative case the data support a modest increase in the derived $t_{\rm delay}$ with decreasing stellar mass; they indicate a {\it lower} limit on $t_{\rm delay}$ of $\sim$ 2 Gyr at  $M_{\rm star}<10^{10.3}M_\odot$.  Thus we consider the main result, or an increasing $t_{\rm delay}$ with decreasing stellar mass, to be quite robust, although quantitatively it is sensitive to having the correct probability distribution function for the photometric redshifts.  We showed in \citet{geec2_data} that, at least for massive galaxies, these probabilities do agree well with the spectroscopic results.  It would be worth repeating these analyses with photometric redshifts from Ultravista; however spectroscopic confirmation will ultimately be required to support our conclusions.

\section{Population properties at infall}\label{sec-app2}
The timescales $t_p$ we derive in \S~\ref{sec-totaltime}, representing the time between accretion and final quenching of star formation, are longer at both $z=0$ and $z=0.9$ than previous estimates from \citet{Wetzel13} and \citet{Mok2}, respectively.  This difference is largely due to assumptions about the galaxy population at infall, which we address here.  There is also extensive discussion of this issue in \citet{Hirschmann}.

\begin{figure}
{\includegraphics[clip=true,trim=0mm 0mm 0mm 0mm,width=3.5in,angle=0]{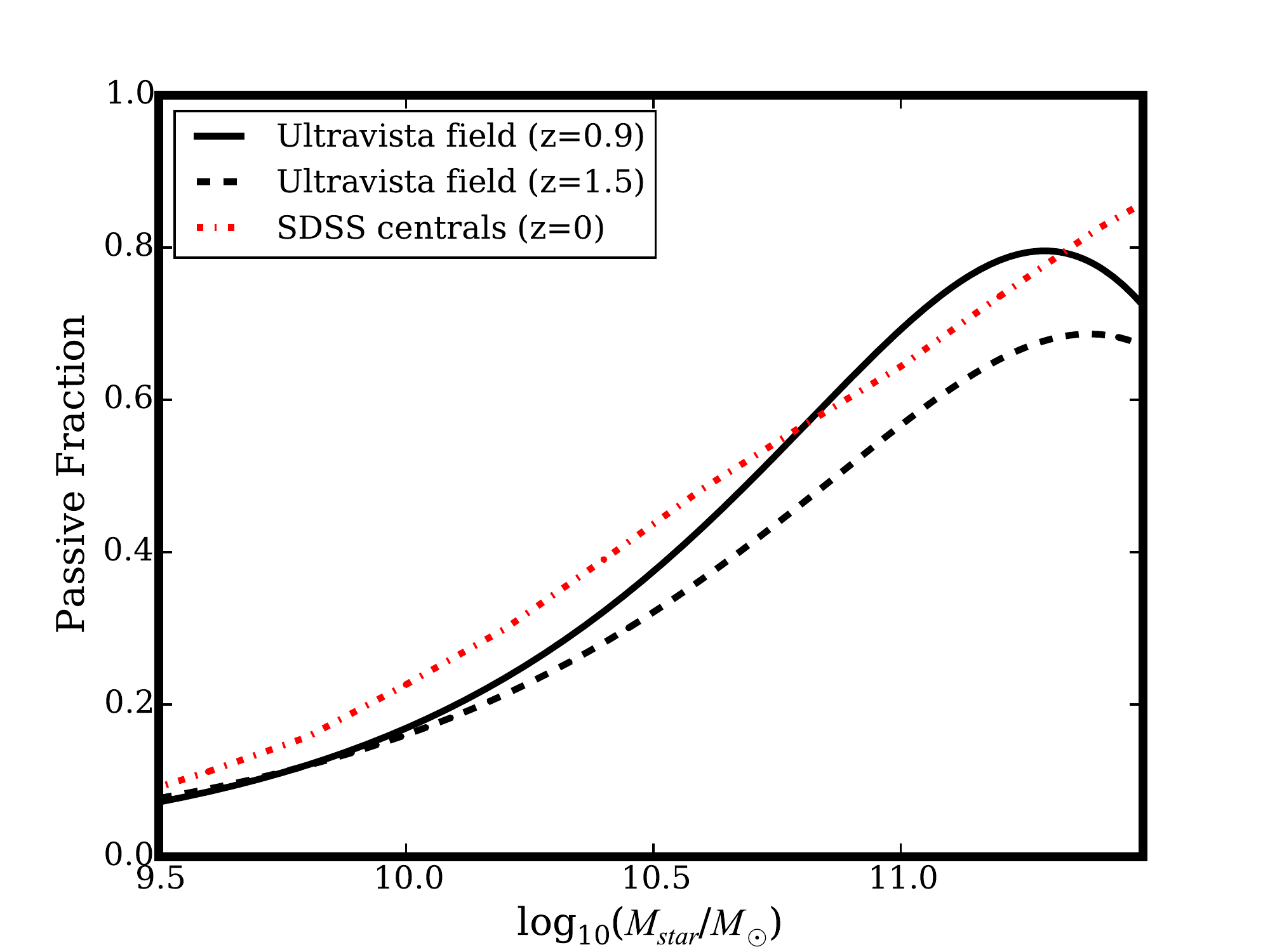}}
        \caption{The passive fraction as a function of stellar mass is shown for central galaxies in SDSS from \citet{Omand}, compared with the field at $z=1.0$ and $z=1.5$ from Ultravista \citep{Ultravista}.  There is little evolution seen in this quantity, which makes our results insensitive to whether we take the properties of galaxies at infall or at the epoch of observation. }
\label{fig:QF_field}
\end{figure}

For the lower--redshift systems, \citet{Wetzel13} derive the relevant timescale from the fraction of galaxies that had star formation quenched while they were satellites, regardless of whether or not they would have quenched as centrals anyway.  This is achieved by comparing the satellite population at the epoch of observation, not to centrals at the same epoch as in Equation~\ref{eq-qe}, but to the predicted properties of the satellite galaxy just prior to accretion.  This makes a large difference, because their model predicts strong evolution in the passive fraction.  However from the Ultravista sample of \citet{Ultravista} and the SDSS analysis from \citet{Omand}, which we use for comparison in this paper, we see in Figure~\ref{fig:QF_field} a much weaker evolution over $0<z<1.5$. Thus, our results for \fc, and the derived $t_{\rm delay}$ timescale, are only weakly affected by this choice.  
In any case, it is not clear which approach, if either, is correct.  By using the infall population as reference, \citet{Wetzel13} are characterizing the net change in satellite properties since infall, regardless of whether or not that change was caused by environmental processes.  By comparing satellites with their contemporary field, as in our approach, we assume that the processes that quench star formation in central galaxies continue unabated after accretion.  The true answer likely lies somewhere in between.

For the GEEC2 sample at $z\sim 0.9$, \citet{Mok2} base their estimate of $t_p$ on the passive fraction itself.  For the spectroscopic sample considered there, this fraction is $\sim 70$ per cent, and from Figure~\ref{fig:McGeeage} this corresponds to $t_p \sim 1$ Gyr, in excellent agreement with the conclusion of \citet{Mok2}.  This shorter timescale results from their assumption that all galaxies are star--forming when they are accreted.  However, \citet{Ultravista} show that $\sim 30$ per cent of galaxies in the relevant mass range were already passive by $z\sim 2$, before most of the mass in these groups was assembled.

\end{document}